\documentclass[PRB,twocolumn,showpacs,amsmath,superscriptaddress]{revtex4-1}
\usepackage[colorlinks,urlcolor=blue,linkcolor=blue,citecolor=blue]{hyperref}
\usepackage{overpic}
\usepackage{subfigure}
\usepackage{dcolumn}
\usepackage{latexsym}
\usepackage{amsfonts}
\usepackage{color}
\usepackage{bm}
\usepackage{array}
\usepackage{booktabs}
\usepackage{caption}
\usepackage{multirow}
\newcommand{\tabincell}[2]{\begin{tabular}{@{}#1@{}}#2\end{tabular}}
\makeatletter

\newcommand{\Rmnum}[1]{\expandafter\@slowromancap\romannumeral #1@}
\makeatother

\begin{document}
	
	\title{Topological Ising pairing states in monolayer and trilayer TaS$_2$}

	\author{Weipeng Chen}
	\affiliation{National Laboratory of Solid State Microstructures and Department of Physics, Nanjing University, Nanjing 210093, China}
	\author{Jin An}
	\affiliation{National Laboratory of Solid State Microstructures and Department of Physics, Nanjing University, Nanjing 210093, China}
	\affiliation{Collaborative Innovation Center of Advanced Microstructures, Nanjing University, Nanjing 210093, China}
	
	\date{\today}
	
	\begin{abstract}
		We study the possibility of topological superconductivity in the noncentrosymmetric monolayer and trilayer TaS$_2$ with out-of-plane mirror symmetry. A gapless time-reversal invariant f+s-wave pairing state with even mirror parity is found to be a promising candidate. This mixing state holds 12(36) nodes at the Fermi pockets around $\Gamma$ for monolayer(trilayer) case and its unconventional superconductivity is consistent with the STM experiments observed in 2$H$-TaS$_2$ thin flakes. Furthermore, with doping or under uniaxial pressure for trilayer 2$H$-TaS$_2$, large-Chern-number time-reversal symmetry breaking mixing states between d+id- and p-ip-wave pairings can be realized in the phase diagram.
	\end{abstract}
	
	\maketitle
	\section{Introduction}\label{sec1}
	Dimensionality plays an important role in superconducting layered materials, which may behave quite differently in two-dimensional(2D) limit. While the layered transition-metal dichalcogenide(TMD) bulk materials exhibit s-wave superconductivity\cite{Clayman1971,Hess1990,Corcoran1994,Boaknin2003,Huang2007}, always coexisting with a competing charge density wave(CDW)\cite{Berthier1976,Mutka1983,Neto2001,Yokoya2001,Guillamon2011,Xi2015N}, their thin flakes have been found to show many new phenomena including the spin-valley locking\cite{Xiao2012,Lu2013,Suzuki2014,Bawden2016,Wu2016,Dey2017}, the valley Hall effect\cite{Xiao2012,Mak2014,Lee2016,Yu2016}, and Ising pairing which is supported by a great enhancement of the in-plane upper critical magnetic field\cite{Lu2015,Xi2015,Saito2016,Xing2017,Lu2018,Sohn2018,Barrera2018}. Recent investigations have also revealed the possibility of unconventional superconductivity in 2D TMDs\cite{Yuan2014,Zhou2016,Sharma2016,Hsu2017,Mockli2018}.
	
	One special member of TMDs family is 2$H$-TaS$_2$, which has attracted much attentions in recent years. In the detached flakes of 2$H$-TaS$_2$, a zero-bias conductance peak(ZBCP) was observed by STM at 0.15 K, indicating unconventional superconductivity\cite{Galvis2014}. This leads to a f-wave pairing scenario in atomic thin 2$H$-TaS$_2$ layers. On the other hand, neither superconductivity nor CDW was observed in the monolayer $H$-TaS$_2$ at 4.7 K\cite{Sanders2016}, which means a strong suppression of CDW in 2D limit of the material. Critical temperature T$_c$ of 2$H$-TaS$_2$ is also found to be strongly enhanced from 0.54 K to about 2.1 K, as the material thickness is decreased from bulk to about five molecular layers\cite{Navarro2016}. Furthermore, T$_c$ of the monolayer could reach 3 K\cite{Barrera2018}. Although the enhancement of T$_{c}$ is believed to result from the suppression of the CDW via dimension reduction\cite{Navarro2016,Yang2018}, what pairing symmetry the superconducting state in 2$H$-TaS$_2$ thin flakes can be and whether it supports topological superconductivity remain unclear.
	
	From the viewpoint of the crystal structure of 2$H$-TaS$_2$, the thin flakes may behave differently for even or odd number of layers, since while 2$H$-TaS$_2$ bulk material always has an inversion center, only the even-layer thin films are still inversion symmetric. Recent experiment has found the difference in gap structure between the trilayer and four-layer 2$H$-NbSe$_2$\cite{Dvir2018}. The odd-layer 2$H$-TaS$_2$ thin films are noncentrosymmetric but always hold out-of-plane mirror symmetry. In this work, we will demonstrate that the Ising superconductivity in monolayer and trilayer TaS$_2$ should be unconventional, and even topologically nontrivial. We find a time-reversal invariant gapless f+s-wave pairing state can be a candidate and furthermore, the time-reversal symmetry(TRS) breaking mixing states between d+id- and p-ip-wave pairings can be induced by doping or uniaxial pressure.

    The paper is organized as follows. In Sec.\ref{sec2}, we introduce our model. In Sec.\ref{sec3}, we analyze the pairing symmetry and classify the gap functions according to crystal symmetry. We also include an appendix to show the technical details of how to make use of the irreducible representations of crystal symmetry to solve the multilayer linearized gap equations. In Sec.\ref{sec4}, we first give the pairing phase diagram, and discuss the Ising pairing states. Then we investigate the topological features of these superconducting states as well as their pairing symmetry transitions with doping or under uniaxial pressure. Finally, we summarize our results in Sec.\ref{sec5}.
	\section{Model}\label{sec2}
\begin{figure}
		\includegraphics[width=7.cm]{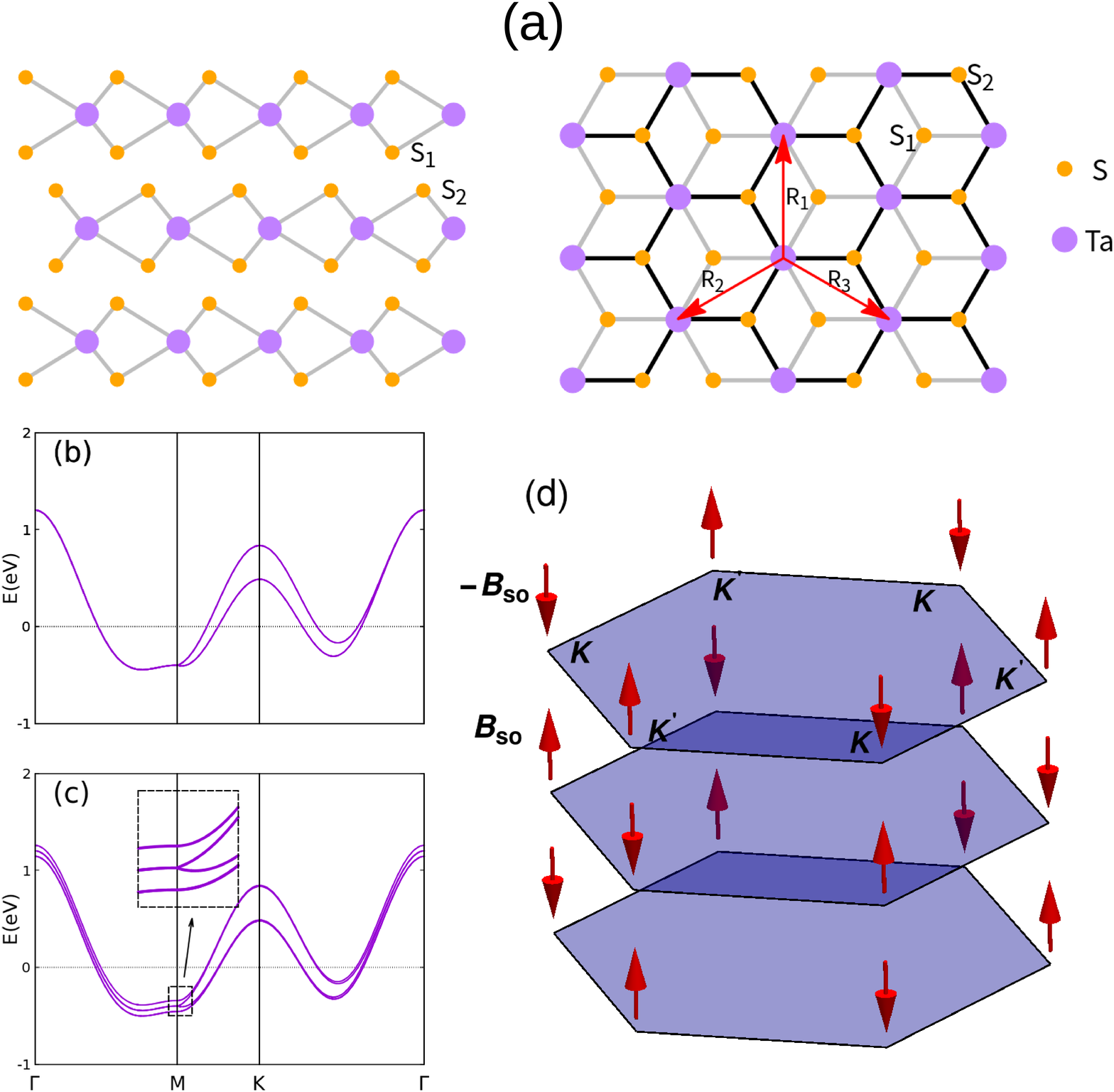}		
		\vspace{-0.0cm}
		\caption{\label{fig01}(color online).(a)The side and top views of atomic structure of trilayer 2$H$-TaS$_2$. Three basis vectors $\bm{R}_j$ are given by the red arrows. The band structures of (b)monolayer $H$-TaS$_2$ and (c)trilayer 2$H$-TaS$_2$. The inset is a blowup of the dotted square near $M$. (d)Spin-valley locking in trilayer 2$H$-TaS$_2$, where the arrows denote the effective magnetic field $B_{so}$ acting on the two valleys $K$ and $K^{'}$ of each layer.}
	\end{figure}

	TMDs have weak van der Waals couplings among TX$_2$($T$=Ta,Nb and $X$=S,Se) layers. An $H$-TX$_2$ monolayer contains three atomic ones each of which forms a triangular lattice, with one $T$ layer sandwiched between two $X$ ones. ``2$H$" in 2$H$-TX$_2$ material means the neighboring TX$_2$ layers in the bulk take the $AbA-CbC$ stacking sequence, preserving the global $D_{6h}$ symmetry. The thin film we study in this paper consists of one or three TaS$_2$ layers. In this and next sections we focus the discussion of model and symmetry analysis on the trilayer case, as shown schematically in Fig. \ref{fig01}(a), since that of the monolayer one is straightforward and parallel. We model the system by simply replacing the three TaS$_2$ layers by three atomic Ta layers in $bbb$ sequence, because superconductivity is believed to occur only within Ta layers. Generically, a thin layer of 2$H$-TaS$_2$ always has Fermi surface(FS) sheets centered at $\Gamma$ and K(K$^\prime$). Although the FS stems respectively from $d_{z^2}$ and $d_{xy/x^2-y^2}$ orbitals\cite{Ge2012,Noat2015,Heil2017}, we simply simulate the bands by a tight-binding model based only on $d_{z^2}$ orbital. Here we ignore CDW as it is believed to be strongly suppressed in the thin 2$H$-TaS$_2$ layers\cite{Navarro2016,Yang2018}. For a trilayer system, the tight-binding Hamiltonian can thus be given by,
	\begin{eqnarray}
	H_0(\bm{k})=
	\begin{bmatrix}
	h_1(\bm{k}) & t_{\perp} & 0 \\
	t_{\perp} & h_2(\bm{k}) & t_{\perp}  \\
	0 & t_{\perp} & h_3(\bm{k})
	\end{bmatrix}
	.\end{eqnarray}
	Here $t_{\perp}$ and $h_l(\bm{k})$ represent the interlayer hopping integral and the intralayer Hamiltonian for layer $l$($l=1,2,3$). The latter can be expressed as
	\begin{eqnarray}
	h_l(\bm{k})=\epsilon (\bm{k})+h^{SO}_{l}(\bm{k})+h^R_{l}(\bm{k})
	,\end{eqnarray}
	where $\epsilon (\bm{k})=-\sum\limits_{j=1}^{3} \{2t_1 \cos k_{j}
	+2t_2 \cos(k_{j}-k_{j+1})\}  -\mu$
	with $k_{j}=\bm{k}\cdot\bm{R_{j}}$, and the unit lattice vectors $\bm{R}_1=\bm{y}$, $\bm{R}_2=-\bm{y}/2-\sqrt{3}\bm{x}/2$, $\bm{R}_3=-\bm{y}/2+\sqrt{3}\bm{x}/2$ and $\bm{R}_4\equiv\bm{R}_1$. Here $\mu$ is the chemical potential, $t_1$, $t_2$ denote the nearest-neighbor(NN) and next-nearest-neighbor(NNN) hoping integrals. We set $(t_1,t_2,t_\perp,\mu)=(-60,-140,-40,0)$ meV to fit the electronic band structure from experiments and DFT calculations \cite{Navarro2016,Sanders2016,Zhao2017,Barrera2018}.
	
	As depicted in Fig. \ref{fig01}(a), the stacking form of sulfur atoms breaks the inversion symmetry in each $H$-TaS$_2$ layer, resulting in the Ising spin-orbit coupling(SOC). Moreover, the middle TaS$_2$ layer can be viewed as being rotated by $180^\circ$ with respect to the top or bottom molecular ones. This leads to a layer-dependent intrinsic SOC given below\cite{Youn2012,Xi2015},
	\begin{eqnarray}
	h^{SO}_{l}(\bm{k})=(-1)^l\beta(\bm{k})\sigma_z,
	\end{eqnarray}
	where $\beta(\bm{k})=\beta_{so}(\sin{k_1}+\sin{k_2}+\sin{k_3})$. For electrons near $K$($K^{'}$), this SOC can be viewed as an effective valley-dependent out-of-plane magnetic field $B_{so}$, favoring spin-up(-down) electrons and alternating among layers, as schematically shown in Fig. \ref{fig01}(d). This gives rise to the well-known spin-valley locking effect\cite{Xiao2012,Lu2013,Suzuki2014,Bawden2016,Wu2016,Dey2017}. The magnitude of $B_{so}$ is estimated to be quite large, about 3000T for TaS$_2$, since the spin splitting at $K$($K'$) is about 348 meV\cite{Sanders2016}, from which $\beta_{so}$ is fixed to be $\beta_{so}=67$ meV. Ising pairing occurs when Cooper pairs are formed between electrons around $K$ and $K^{'}$, spin oriented oppositely but along the out-of-plane direction. Ising superconductivity in TMDs has been supported experimentally by the significant enhancement of the in-plane upper critical magnetic field\cite{Lu2015,Xi2015,Saito2016,Xing2017,Liu2017,Lu2018,Sohn2018,Barrera2018}.

    Rashba SOC is also taken into account. It takes the following form,
	\begin{eqnarray}
	h^R_{l}(\bm{k})=\alpha^{l}\bm{g(k)}\cdot \bm{\sigma}
	,\end{eqnarray}
	where $\bm{g}(\bm{k})=(-Im [\lambda(\bm{k})],-Re[\lambda(\bm{k})],0)$, with $\lambda (\bm{k})=\sin{k_1}+e^{-i\frac{2\pi}{3}}\sin{k_3}+e^{-i\frac{4\pi}{3}}\sin{k_3}$. Since the Rashba SOC is strongly dependent on the surface normal, the coupling constants should be layer-dependent, which are assumed to be $(\alpha^{1},\alpha^{2},\alpha^{3})=(\alpha_R,0,-\alpha_R)$ for a free-standing trilayer 2$H$-TaS$_2$, with $\alpha_R$ the coupling strength. This kind of coupling also appears in the layered system with superlattice structures\cite{Mizukami2011,Goh2012,Shimozawa2014}. The Rashba SOC favors electrons' spin oriented along the in-plane directions, which means it will compete with the Ising SOC and weaken the effect of spin-valley locking and hence Ising pairing. As long as $\alpha_R\ll\beta_{so}$, Ising pairing is expected to be still dominant.

    The band structures for monolayer $H$-TaS$_2$ and trilayer 2$H$-TaS$_2$ without Rashba SOC are shown in Fig. \ref{fig01}(b) and (c), respectively. For monolayer $H$-TaS$_2$, the band is doubly degenerate at $\Gamma$ but spin split at $K$($K'$), as expected. For trilayer 2$H$-TaS$_2$, there are four bands in total but two of the four are doubly degenerate due to accidental degeneracy. The degeneracy can be lifted when the middle layer takes a slightly different value of $\beta_{so}$ to the top/bottom layer, which still preserves the crystal symmetry of trilayer 2$H$-TaS$_2$.

	\section{Pairing Interaction and Symmetry Analysis}\label{sec3}
	The trilayer 2$H$-TaS$_2$ system belongs to crystal $D_{3h}$, while each TaS$_2$ monolayer(especially the top/bottom layer) only has $C_{3v}$ symmetry. Since no pairing occurs between layers, the pairing basis gap functions $\Delta^{\Gamma,\alpha}_{i}(\bm{k})$ can thus be classified according to the irreducible representations(IRs) of $C_{3v}$\cite{Yuan2014}, as shown in Table \ref{tab1}. According to whether the total spin $S$ of the Cooper pair is $1$ or $0$, $\Delta^{\Gamma,\alpha}_{i}(\bm{k})$ takes the following two different kinds of forms:
	\begin{eqnarray}\Delta^{\Gamma,\alpha}_{i}(\bm{k})=
	\begin{cases}
	\Psi^{\Gamma,\alpha}_{i} i\sigma_y & {singlet \ pairing}\\
	\bm{d}^{\Gamma,\alpha}_{i} \cdot \bm{\sigma} i \sigma_y & {triplet \ pairing}
	\end{cases}
	,\end{eqnarray}
	where $\Psi^{\Gamma,\alpha}_{i}$ is the singlet order parameter while $\bm{d}^{\Gamma,\alpha}_{i}$ denotes the director for the triplet pairing. Here $\Gamma$ represents the IR of $C_{3v}$ symmetry with dimension $d^{\Gamma}$ and it can be $A_{1}(d^{A_{1}}=1)$, $A_{2}(d^{A_{2}}=1)$ or $E(d^{E}=2)$. The index $\alpha$ is used to distinguish different types of representations in the same IR $\Gamma$, and $i$ runs from $1$ to $d^{\Gamma}$. As an example, for $\Gamma=A_1$, $\alpha$ can be `on', `nn', `z' and `xy'. All these $\Delta^{\Gamma,\alpha}_{i}(\bm{k})$ are so normalized $\frac{2}{N}\sum_{\bm{k}}\mid\Psi^{\Gamma,\alpha}_{i}\mid^{2}=1$(or $\frac{2}{N}\sum_{\bm{k}}\mid\bm{d}^{\Gamma,\alpha}_{i}\mid^{2}=1$) as to satisfy the following orthogonal relations between IRs:
	\begin{eqnarray}\label{eqn6}
	\frac{1}{N}\sum_{\bm{k}} Tr\{\Delta^{ \dagger \Gamma^\prime,\alpha}_{i}(\bm{k}) \Delta^{\Gamma,\beta}_{j}(\bm{k})\}=\delta_{\Gamma^\prime \Gamma}\delta_{\alpha \beta}\delta_{ij}
	,\end{eqnarray}
	where N is the total number of unit cells in the system. See Appendix for the detail.

    Although all the basis gap functions in Table \ref{tab1} can be realized in principle in materials with crystal $C_{3v}$ symmetry, the in-plane triplet pairing components such as $\bm{d}^{A_1,xy}$, $\bm{d}^{A_2,xy}$, $\bm{d}^{E,xy}$, $\bm{d}^{E,\widetilde{xy}}$ actually never occur as long as Ising pairing is dominant in 2$H$-TaS$_2$, which is guaranteed by the strong intrinsic SOC. This is actually also confirmed by our numerical calculations in Sec.\ref{sec4}.
	
	Without considering the pairing mechanism, the pairing term for a multilayer 2$H$-TaS$_2$ can be generically written as:
	\begin{eqnarray}
	H_{int}&=&
	\frac{1}{2}\sum V_{s_1s_2s_3s_4}(\bm{k},\bm{k}^\prime)
	c^{\dagger}_{l,\bm{k}s_1}c^{\dagger}_{l,-\bm{k}s_2}\nonumber\\
	&&\times c_{l,\bm{k}^{\prime}s_3}c_{l,-\bm{k}^{\prime}s_4}
	,\end{eqnarray}
	where the sum runs over the repeated indices. $V_{s_1s_2s_3s_4}(\bm{k},\bm{k}^\prime)$ is the pairing interaction and $c^{\dagger}_{l,\bm{k}s}$ the creation operator for an electron with spin $s$ at layer $l$. According to $C_{3v}$ symmetry and taking the on-site and NN parings into account, $V_{s_1s_2s_3s_4}(\bm{k},\bm{k}^\prime)$ can be expanded as follows,
	
	\begin{eqnarray}
	&&V_{s_1s_2s_3s_4}(\bm{k},\bm{k}^\prime) \nonumber \\
	=&&v_0\Psi^{A_{1},on}\Psi^{*A_{1},on}(i\sigma_y)_{s_1s_2}(i\sigma_y)_{s_3s_4}+  \nonumber \\
	&&v_1 \sum_{\Gamma,i}\Psi^{\Gamma,nn}_{i} \Psi^{*\Gamma,nn}_{i}(i\sigma_y)_{s_1s_2}(i\sigma_y)_{s_3s_4}+  \nonumber \\
	&&v_1 \sum_{\Gamma,\alpha,i} [\bm{d}^{\Gamma,\alpha}_{i}\cdot \bm{\sigma}i\sigma_y]_{s_1s_2}[\bm{d}^{\Gamma,\alpha}_{i}\cdot \bm{\sigma}i\sigma_y]^*_{s_3s_4}
	,\end{eqnarray}
	where $v_0(v_1)$ is the pairing strength for the on-site(NN) pairing interaction.
	
	\renewcommand\arraystretch{1.5}
	\begin{table}[!hbp]
		\centering
		\caption{Classification of spin-singlet and spin-triplet basis gap functions based on $C_{3v}$ symmetry, where $C(\bm{k})=\frac{1}{\sqrt{3}}(\cos{k_1}+\cos{k_2}+\cos{k_3}), C_+ (\bm{k})=\frac{1}{\sqrt{3}}(\cos{k_1}+e^{i\frac{2\pi}{3}}\cos{k_2}+e^{i\frac{4\pi}{3}}\cos{k_3}), S(\bm{k})=\frac{1}{\sqrt{3}}(\sin{k_1}+\sin{k_2}+\sin{k_3}), S_+ (\bm{k})=\frac{1}{\sqrt{3}}(\sin{k_1}+e^{i\frac{2\pi}{3}}\sin{k_2}+e^{i\frac{4\pi}{3}}\sin{k_3}), C_- (\bm{k})=C_+^* (\bm{k}), S_- (\bm{k})=S_+^* (\bm{k})$, with $\bm{x}_\pm=(\bm{x}\pm i\bm{y})/2$.}
		\begin{tabular}{cccc}
			\hline
			\specialrule{0.05em}{3pt}{3pt}
			$\Gamma$ && Singlet & Triplet  \\
			\specialrule{0.05em}{3pt}{3pt}
			$A_1$ && \tabincell{c}{$\Psi^{A_1,on}_{}=\frac{1}{\sqrt{2}}$ \\  $\Psi^{A_1,nn}_{}=C(\bm{k})$} & \tabincell{c}{$\bm{d}^{A_1,z}_{}=S(\bm{k})\bm{z}$ \\$\bm{d}^{A_1,xy}_{}=i[S_-(\bm{k})\bm{x}_+-S_+(\bm{k})\bm{x}_-$]}\\
			\specialrule{0.05em}{3pt}{3pt}
			$A_2$ && & $\bm{d}^{A_2,xy}_{}=S_-(\bm{k})\bm{x}_++S_+(\bm{k})\bm{x}_-$\\
			\specialrule{0.05em}{3pt}{3pt}
			E & &  \tabincell{c}{$\begin{cases} \Psi^{E,nn}_{1}=C_+(\bm{k})    \\ \Psi^{E,nn}_{2}=C_-(\bm{k})\end{cases}$} & \tabincell{c}{$\begin{cases} \bm{d}^{E,z}_{1}=S_+(\bm{k})\bm{z}\\ \bm{d}^{E,z}_{2}=S_-(\bm{k})\bm{z}\end{cases} $  \\$\begin{cases}\bm{d}^{E,xy}_{1}=\sqrt{2}S(\bm{k})\bm{x}_+ \\
				\bm{d}^{E,xy}_{2}=-\sqrt{2}S(\bm{k})\bm{x}_-\end{cases}$  \\$\begin{cases}\bm{d}^{E,\widetilde {xy}}_{1}=\sqrt{2}S_-(\bm{k})\bm{x}_-\\
				\bm{d}^{E,\widetilde {xy}}_{2}=-\sqrt{2}S_+(\bm{k})\bm{x}_+\end{cases}$}\\
			\specialrule{0.05em}{3pt}{3pt}
			\hline
		\end{tabular}
		\label{tab1}
	\end{table}

	Because the trilayer 2$H$-TaS$_2$ system lacks an inversion center, generically there exists mixing between singlet and triplet pairings. The corresponding $6\times6$ pairing gap function $\Delta(\bm{k})$ for the superconducting trilayer system can thus be expanded in the basis of $\Delta^{\Gamma,\alpha}_{i}(\bm{k})$ as,
	\begin{eqnarray}\label{eqn9}
	\Delta(\bm{k})/\Delta&\equiv&
	\begin{bmatrix}
	\Delta_{l=1}(\bm{k})&& \\
	&\Delta_{l=2}(\bm{k})& \\
	&&\Delta_{l=3}(\bm{k}) \\
	\end{bmatrix} \nonumber
	\\
	&=&\sum\limits_{\alpha,i}
	b^{\Gamma,\alpha}_{i} \begin{bmatrix}
	\chi^{\Gamma,\alpha}_{l=1,i} && \\
	&\chi^{\Gamma,\alpha}_{l=2,i}& \\
	&&\chi^{\Gamma,\alpha}_{l=3,i}\\
	\end{bmatrix}\Delta^{\Gamma,\alpha}_{i}(\bm{k})
	,\end{eqnarray}
	where $\Delta$ is the gap value, and the expansion is made only for a definite IR $\Gamma$, as the mixing among different IRs is forbidden\cite{Sigrist1991}. Here $\Delta_{l}(\bm{k})$ is the $2\times2$ gap matrix for layer $l$. The expansion coefficients $b^{\Gamma,\alpha}_{i}$ can be normalized as $\sum \limits_{\alpha,i} |b^{\Gamma,\alpha}_{i}|^{2}=1$, while the relative paring amplitudes $\chi_{l,i}^{\Gamma,\alpha}$ are always normalized as $\sum \limits_l |\chi_{l,i}^{\Gamma,\alpha}|^{2}=1$. When the most energetically favorable pairing gap function $\Delta(\bm{k})$ is found, the BdG Hamiltonian of this superconducting trilayer system takes the standard form,
	\begin{eqnarray}
	H_{BdG}(\bm{k})=
	\begin{bmatrix}
	H_0(\bm{k}) & \Delta (\bm{k}) \\
	\Delta^\dagger (\bm{k}) & -H_0^T(\bm{-k})
	\end{bmatrix}
	.\end{eqnarray}

   This $D_{3h}$ trilayer system also has an out-of-plane mirror symmetry, with the mirror plane lying in the middle Ta layer. For the normal state, we have $[M_{xy},H_0(\bm{k})]=0$, with
	\begin{eqnarray}
	M_{xy}=
	\begin{bmatrix}
	0  & 0 & 1\\
	0  & 1 & 0\\
	1  & 0 & 0\\
	\end{bmatrix}
	i\sigma_z
	,\end{eqnarray}
	which exchanges the top layer with bottom layer and reverses the in-plane spins. For the superconducting state, the mirror symmetry requires $[M^{\pm},H_{BdG}(\bm{k})]=0$, with mirror operators $M^{\pm}=\left(
	\begin{array}{cc}
	M_{xy}& 0\\
	0& \pm M_{xy}^* \\
	\end{array}
	\right)$, leading to a requirement that
	\begin{eqnarray}
	M_{xy}\Delta(\bm{k})M_{xy}^T=\pm \Delta(\bm{k}),
	\end{eqnarray}
	where the symbol ``+(-)'' represents even(odd) mirror parity\cite{Yoshida2015}. Therefore in the trilayer 2$H$-TaS$_2$ system, for an even-mirror-parity state, its singlet components or triplet ones with $\bm{d}^{\Gamma}\parallel\bm{z}$, and for an odd-mirror-parity state, its triplet components with $\bm{d}^{\Gamma}\perp\bm{z}$, satisfy,
	\begin{eqnarray}
	\chi^{\Gamma,\alpha}_{l=1,i}(\bm{k})=\chi^{\Gamma,\alpha}_{l=3,i}(\bm{k}).
	\end{eqnarray} For an odd-mirror-parity state, its singlet components or triplet ones with $\bm{d}^{\Gamma}\parallel\bm{z}$, and for an even-mirror-parity state, its triplet components with $\bm{d}^{\Gamma}\perp\bm{z}$, satisfy instead the requirements,
	\begin{eqnarray}
	\begin{cases}
	\chi^{\Gamma,\alpha}_{l=1,i}(\bm{k})=-\chi^{\Gamma,\alpha}_{l=3,i}(\bm{k})=\frac{1}{\sqrt{2}} \\
	\chi^{\Gamma,\alpha}_{l=2,i}(\bm{k})=0
	\end{cases}
	.\end{eqnarray} These symmetry properties are summarized in Table \ref{tab2}.
	
		\renewcommand\arraystretch{1}
		
		\renewcommand\arraystretch{1.5}
		\begin{table}[b]
			\centering
			\caption{Symmetry of relative pairing amplitudes $\bm{\chi}^{\Gamma,\alpha}_{i}$ for a pairing gap function of the trilayer 2$H$-TaS$_2$, classified by the out-of-plane mirror symmetry.}
			\begin{tabular}{p{1cm}<{\centering}p{3cm}<{\centering} p{4cm}<{\centering}}
				\hline
				\specialrule{0.05em}{3pt}{3pt}
				Mirror Parity&Components&$\bm{\chi}^{\Gamma,\alpha}_{i}\equiv (\chi^{\Gamma,\alpha}_{1,i},\chi^{\Gamma,\alpha}_{2,i},\chi^{\Gamma,\alpha}_{3,i})$ \\
				\specialrule{0.05em}{3pt}{3pt}
				Even &\tabincell{c}{$\psi^{\Gamma,\alpha}_{i}$ or $\bm{d}_{i}^{\Gamma,\alpha} || \bm{z}$ \\ $\bm{d}_{i}^{\Gamma,\alpha} \perp \bm{z}$} &\tabincell{c}{$\chi^{\Gamma,\alpha}_{1,i}=\chi^{\Gamma,\alpha}_{3,i}$ \\ ($\frac{1}{\sqrt{2}}$,  0, -$\frac{1}{\sqrt{2}})$}\\
				\specialrule{0.05em}{3pt}{3pt}
				Odd &\tabincell{c}{$\psi^{\Gamma,\alpha}_{i}$ or $\bm{d}_{i}^{\Gamma,\alpha} || \bm{z}$ \\ $\bm{d}_{i}^{\Gamma,\alpha} \perp \bm{z}$} &\tabincell{c}{($\frac{1}{\sqrt{2}}$, 0, -$\frac{1}{\sqrt{2}}$) \\ $\chi^{\Gamma,\alpha}_{1,i}=\chi^{\Gamma,\alpha}_{3,i}$}\\
				\specialrule{0.05em}{3pt}{3pt}
				\hline
			\end{tabular}
			\label{tab2}
		\end{table}
		\renewcommand\arraystretch{1}
	
	In order to determine the detailed paring symmetry $\Delta(\bm{k})$ and T$_c$ of the superconducting state, we solve the following coupled linearized gap equations,
	\begin{eqnarray}\label{eqn15}
	[\Delta_{l}(\bm{k})]_{s_1s_2}&=& \frac{-T_c}{N}\sum\limits_{\omega_{n}\bm{k}^\prime s_3 s_4}V_{s_1s_2s_3s_4}(\bm{k},\bm{k}^\prime) \nonumber\\
	&& \times [G(k)\Delta(\bm{k}^\prime)G^{\tau}(-k^\prime)]_{\substack{l,l \\s_3s_4}}
	,\end{eqnarray}\label{eq15}
	where $G(k)=[i\omega_n-H_0(\bm{k})]^{-1}$ with $k\equiv(\bm{k},i\omega_{n})$ is the Matsubara Green's function for the normal state. By solving the above eigenequation for each IR $\Gamma$, one obtain its eigenstates and corresponding T$_c$. The most favorable pairing state just corresponds to the eigenstate with the highest T$_c$.  See Appendix for the detail.

	\begin{figure}[t]
		\includegraphics[width=7.5cm]{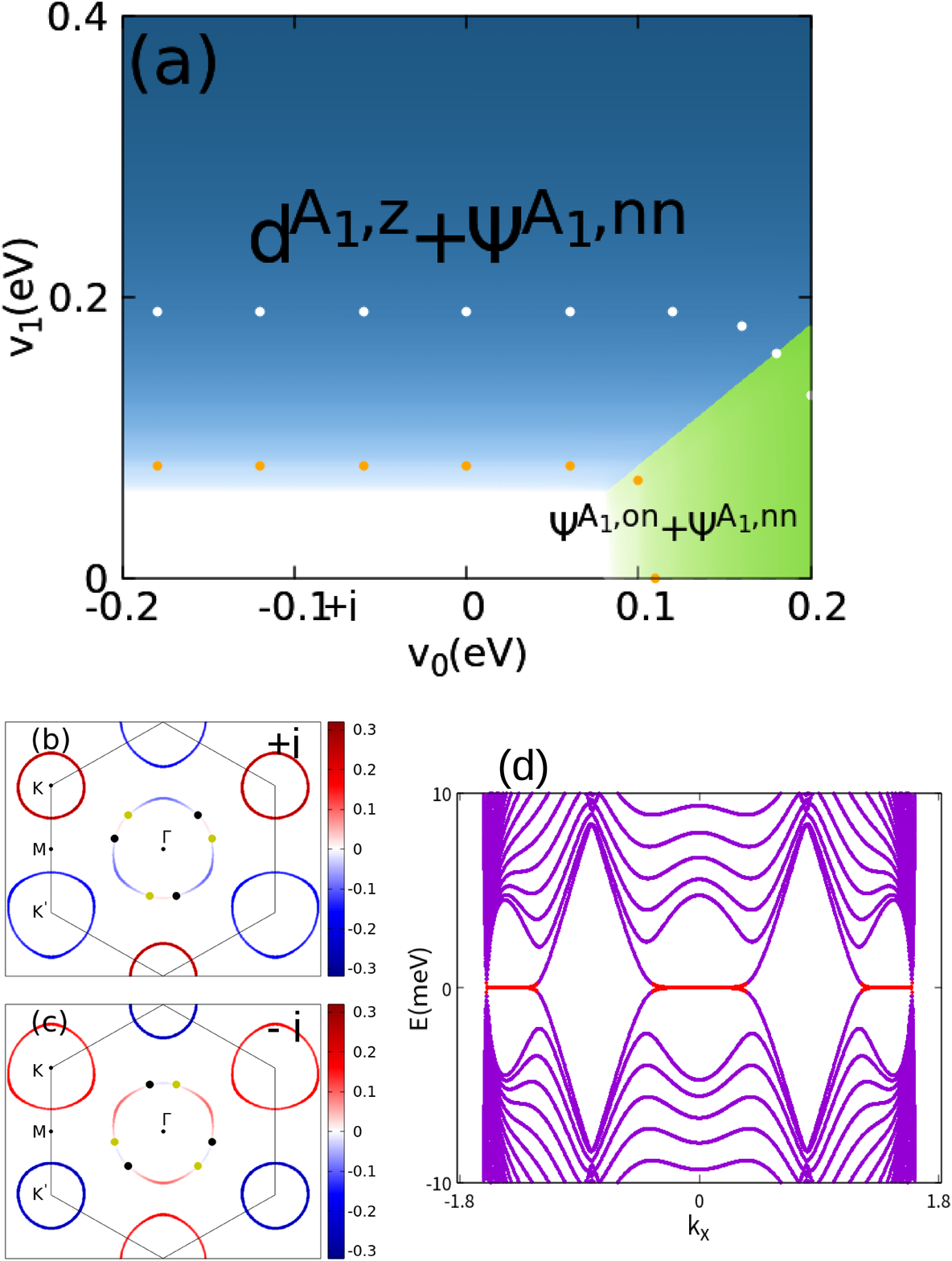}		
		\vspace{-0.cm}
		\caption{\label{fig02}(color online).
			(a) $v_0$ versus $v_1$ pairing phase diagram of monolayer $H$-TaS$_2$, where the white(orange) dots show the equal T$_c$ line with T$_c$=10 K(1$\times 10^{-2}$ K). The FS and nodes of (b) sector $+i$ and (c) sector $-i$ for the $\bm{d}^{A_1,z}+\Psi^{A_1,nn}$ state. The FS are colored in red(blue) to represent the positive(negative) gap functions. The black(yellow) dots denote the nodes with winding number +1(-1). Parameters are chosen to be $(v_{0},v_{1})=(0.1,0.16)$ eV, $(\Delta^s,\Delta^f)=\Delta(-0.252,0.968)$ with $\Delta=1$ K. (d)The corresponding edge band for sector $+i$ with open-boundary conditions along y direction, where for better view, $\Delta$ has been enlarged 20 times without changing the topological nature of the system.
		}
	\end{figure}
	
	\section{Result and discussion}\label{sec4}
	
	\subsection{Monolayer $H$-TaS$_2$ and trilayer 2$H$-TaS$_2$}
	
	As a comparison, we first study the superconducting state for the monolayer $H$-TaS$_2$ without Rashba SOC. This monolayer system preserves the out-of-plane mirror symmetry $i\sigma_{z}$ which has a requirement on the directors $\bm{d}^\Gamma$ of the triplet components: $\bm{d}^\Gamma$ is either parallel or normal to the TaS$_2$ plane, as shown in Table \ref{tab1}. When $\bm{d}^\Gamma\parallel\bm{z}$($\bm{d}^\Gamma\perp\bm{z}$), the mirror parity of the superconducting state is even(odd). Thus in this noncentrosymmetric monolayer system, the singlet components can be mixing merely with the triplet components having $\bm{d}^\Gamma\parallel\bm{z}$ to form an even-mirror-parity state, while the odd-mirror-parity state is a pure triplet state only consisting of components with $\bm{d}^\Gamma\perp\bm{z}$. The odd-mirror-parity state can be ruled out as Ising pairing is dominant due to strong Ising SOC. Fig. \ref{fig02}(a) is the pairing phase diagram calculated for the monolayer $H$-TaS$_2$. One finds two even-mirror-parity states, both of which are mixing ones: the mixing state between $\bm{d}^{A_1,z}$ and $\Psi^{A_1,nn}$, and that between $\Psi^{A_1,on}$ and $\Psi^{A_1,nn}$. The latter is a topologically trivial s-wave state, which is fully gapped on all the FS, while the former could be topologically nontrivial, since $\bm{d}^{A_1,z}$ is a $S_z=0$ f-wave pairing, which holds three nodal lines.
	
\begin{figure}[t]
	\centering
	\includegraphics[width=7.5cm]{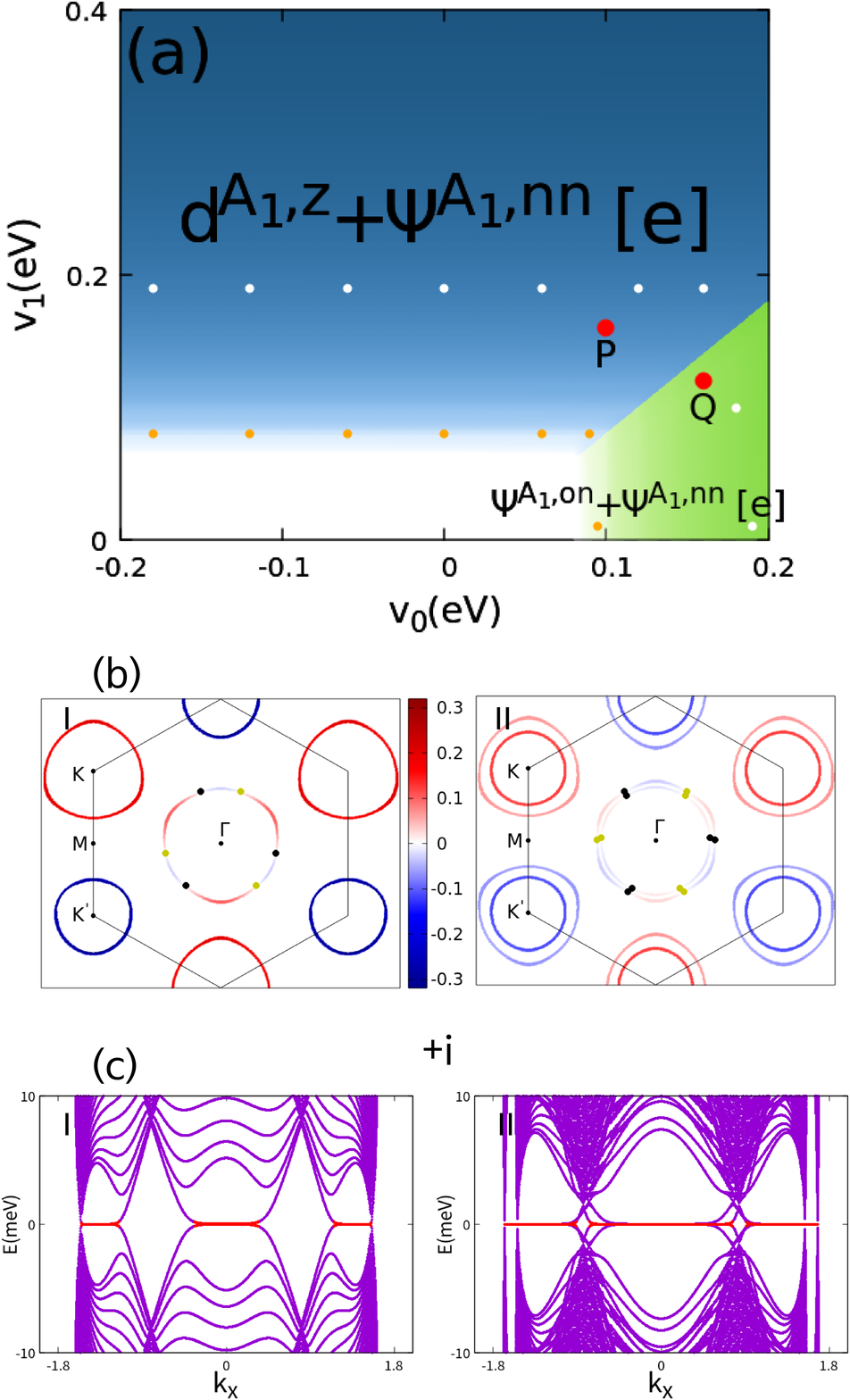}	
	\vspace{-0.cm}
	\caption{(color online).\label{fig03}(a)The same as Fig. \ref{fig02}, except that it is for trilayer 2$H$-TaS$_2$. Symbol `e' represents even mirror parity. (b)The FS and nodes of of sector $+i$ for the representative mixing state denoted by `P' in (a), shown respectively for the two subsectors. Here the gap value is chosen to be $\Delta \approx 1$ K. (c) The corresponding edge band for sector $+i$ .}
\end{figure}

		\renewcommand\arraystretch{1.5}
		\begin{table*}
			\caption{The detailed expansion coefficients for the four representative mixing pairing states of IR $A_{1}$ denoted by `P', `Q' in Fig. \ref{fig03}(a) and `R', `T' in Fig. \ref{fig04}(a).}
			\begin{tabular}{|p{4cm}<{\centering}|p{4cm}<{\centering}|p{4cm}<{\centering}|p{4cm}<{\centering}|c|}
				\hline
				\hline
				$b^{A_1,on}$; $\bm{\chi}^{A_1,on}$ & $b^{A_1,nn}$; $\bm{\chi}^{A_1,nn}$&  $b^{A_1,z}$; $\bm{\chi}^{A_1,z}$ &$b^{A_1,xy}$; $\bm{\chi}^{A_1,xy}$&State\\
				\hline
				&     0.179; (-0.642, 0.419,-0.642)&0.983; (0.586, 0.560, 0.586)&&$P$\\
				0.990; (0.572, 0.584, 0.572)&-0.044; (0.570, 0.580, 0.570) &&&$Q$\\
				&     0.151; (-0.627, 0.455,-0.627)&0.892; (0.583, 0.566, 0.583)&-0.424(0.707, 0,-0.707)&$R$\\
				0.967; (0.595, 0.539, 0.595)&0.100; (-0.570,-0.580,-0.570) & &-0.225(0.707, 0,-0.707)&$T$\\
				\hline
				
			\end{tabular}
			\label{tab3}
		\end{table*}
		\renewcommand\arraystretch{1}

	To investigate the topological feature of the $\bm{d}^{A_1,z}+\Psi^{A_1,nn}$ state, which can be favorable when the NN paring is dominant, one can block-diagonalize the monolayer Hamiltonian into $\pm$i(mirror parity) sectors according to the mirror symmetry,
	$ H_{BdG}(\bm{k}) \rightarrow \begin{bmatrix}
	H_{+i}(\bm{k}) &0 \\
	0 &H_{-i}(\bm{k}) \\
	\end{bmatrix}$. The two Hamiltonians are connected with each other by TRS: $H_{-i}(\bm{k})=H_{+i}(-\bm{k})$.
	The excitation spectrum is: $E_{+i}(\bm{k})=\sqrt{(\epsilon(\bm{k})-\beta(\bm{k}))^2+(\Delta^f(\bm{k})+\Delta^s(\bm{k}))^2}$, $E_{-i}(\bm{k})=E_{+i}(-\bm{k})$, with $\Delta^f(\bm{k})=\Delta^f S(\bm{k})$ and $\Delta^s(\bm{k})=\Delta^s C(\bm{k})$. Here $(\Delta^s,\Delta^f)=\Delta(b^{A_1,nn},b^{A_1,z})$, which are found to take real values. There exist 12 nodal points and all nodes are located at the Fermi pockets around $\Gamma$, while the superconducting state is fully gapped with its order parameter taking opposite signs at the Fermi pockets around $K$ and $K'$(see Fig. \ref{fig02}(b)-(c)). Because the superconducting state is chiral symmetric, each node is characterized by a winding number(WN). It can be calculated via,
	\begin{eqnarray}\label{eqn17}
	\nu_{\pm i}=\frac{1}{2 \pi i}\oint dk Tr\partial_{k} lnq_{\pm i}(\bm{k})
	,\end{eqnarray}
	where the integration is along any small loop around the node and $q_{\pm i}(\bm{k})=i[\Delta^{f}(\bm{k})\pm\Delta^{s}(\bm{k})]+[\epsilon(\bm{k})\mp\beta(\bm{k})]$ appears in the off-diagonal representation of $H_{\pm i}(\bm{\bm{k}})$: $H_{\pm i}(\bm{k})\rightarrow
	\begin{bmatrix}
	0 & q_{\pm i}(\bm{k}) \\
	q^\dagger_{\pm i}(\bm{k}) & 0
	\end{bmatrix}$. The edge states calculated for $H_{+i}$ with zigzag boundary shown in Fig. \ref{fig02}(d) give the Majorana flat edge band lying between the projections of pairs of nodes having the opposite WNs.

	Now we turn to study the superconducting state in trilayer 2$H$-TaS$_2$. Its pairing phase diagram is illustrated in Fig. \ref{fig03}(a), which is similar to that of the monolayer $H$-TaS$_2$: There are two mixing Ising paring states, including a trivial full-gap s-wave state and a topologically nontrivial mixing one between $\bm{d}^{A_{1},z}$ and $\Psi^{A_{1},nn}$. The $\bm{d}^{A_{1},z}+\Psi^{A_{1},nn}$ state is dominated by the f-wave triplet component $\bm{d}^{A_{1},z}$, and the expansion coefficients for a representative state in the phase diagram is given in Table \ref{tab3}. As before, we block-diagonalize the mirror-symmetric $H_{BdG}(\bm{k})$ into two TRS connected sectors $H_{\pm i}(\bm{k})$. Due to $S_{z}$ conservation, $H_{+i}(\bm{k})$ can be further reduced by a basis change to two smaller sectors $\Rmnum{1}$ and $\Rmnum{2}$:
	$H_{+i}(\bm{k})\rightarrow\begin{bmatrix}
	H_{+i}^{\Rmnum{1}}(\bm{k})& 0 \\
	0 &H_{+i}^{\Rmnum{2}}(\bm{k})\\
	\end{bmatrix}$ with
	\begin{widetext}
		\begin{eqnarray}
		H_{+i}^{\Rmnum{1}}(\bm{k})=
		\begin{bmatrix}
		\epsilon(\bm{k})+\beta(\bm{k}) & \Delta_1^f(\bm{k})-\Delta_1^s(\bm{k}) \\
		\Delta_1^f(\bm{k})-\Delta_1^s(\bm{k}) &	-\epsilon(\bm{k})-\beta(\bm{k}) \\
		\end{bmatrix}
		,\end{eqnarray}
		\begin{eqnarray}
		H_{+i}^{\Rmnum{2}}(\bm{k})=
		\begin{bmatrix}
		\epsilon(\bm{k})-\beta(\bm{k})  & \sqrt{2} t_\perp& 0& \Delta_2^f(\bm{k})+\Delta_2^s(\bm{k}) \\
		\sqrt{2} t_\perp&\epsilon(\bm{k})+\beta(\bm{k}) &  \Delta_1^f(\bm{k})+\Delta_1^s(\bm{k}) & 0 \\
		0 & \Delta_1^f(\bm{k})+\Delta_1^s(\bm{k})&  -\epsilon(\bm{k})+\beta(\bm{k})& -\sqrt{2} t_\perp \\
		\Delta_2^f(\bm{k})+\Delta_2^s(\bm{k}) &0 & -\sqrt{2} t_\perp& -\epsilon(\bm{k})-\beta(\bm{k}) \\
		\end{bmatrix},
		\end{eqnarray}
	\end{widetext}
	Here $H_{\pm i}^{\Rmnum{1}(\Rmnum{2})}(\bm{k})$ is written in the basis of ($c_{-,\bm{k}\downarrow},c^{\dagger}_{-,\bm{-k}\uparrow}$) and ($c_{+,\bm{k}\uparrow},c_{2,\bm{k}\uparrow},c^{\dagger}_{+,-\bm{k}\downarrow},c^{\dagger}_{2,-\bm{k}\downarrow}$), respectively, with $c_{\pm,\bm{k}\sigma}=\frac{1}{\sqrt{2}}(c_{1,\bm{k}\sigma}\pm c_{3,\bm{k}\sigma})$. For a definite band $n$, the superconducting state on its FS can be described by the effective gap function, or the projected gap, which is given by the diagonal entry $\widetilde{\Delta}_{n,n}(\bm{k})$, where the effective gap function $\widetilde{\Delta} (\bm{k})$ is given by $\widetilde{\Delta} (\bm{k})=U^\dagger(\bm{k})\Delta (\bm{k})U^*(-\bm{k})$. Here the unitary matrix $U(\bm{k})$ diagonalizes $H_{0}(\bm{k})$: $U^{\dagger}(\bm{k})H_{0}(\bm{k})U(\bm{k})=D(\bm{k})$, with $D(\bm{k})$ a diagonal matrix. At the FS around $K$ or $K'$, the superconducting state is fully gapped and the projected gap takes positive and negative values alternatively, as shown in Fig. \ref{fig03}(b). The same reason leads to the existence of nodes at the FS around $\Gamma$. Around each FS, the sign of the projected gap changes 6 times and there are 6(12) nodes for $H^{\Rmnum{1}}_{+i}(\bm{k})$($H^{\Rmnum{2}}_{+i}(\bm{k}))$ and 36 nodes in total for $H_{BdG}(\bm{k})$. Since both $H^{\Rmnum{1}}_{+i}(\bm{k})$ and $H^{\Rmnum{2}}_{+i}(\bm{k})$ hold chiral symmetry, each node has a well-defined WN $\pm1$ which can still be calculated via Eq.(\ref{eqn17}), as exhibited in Fig. \ref{fig03}(b).
	\begin{figure}[htp]
		\centering
		\includegraphics[width=7.5cm]{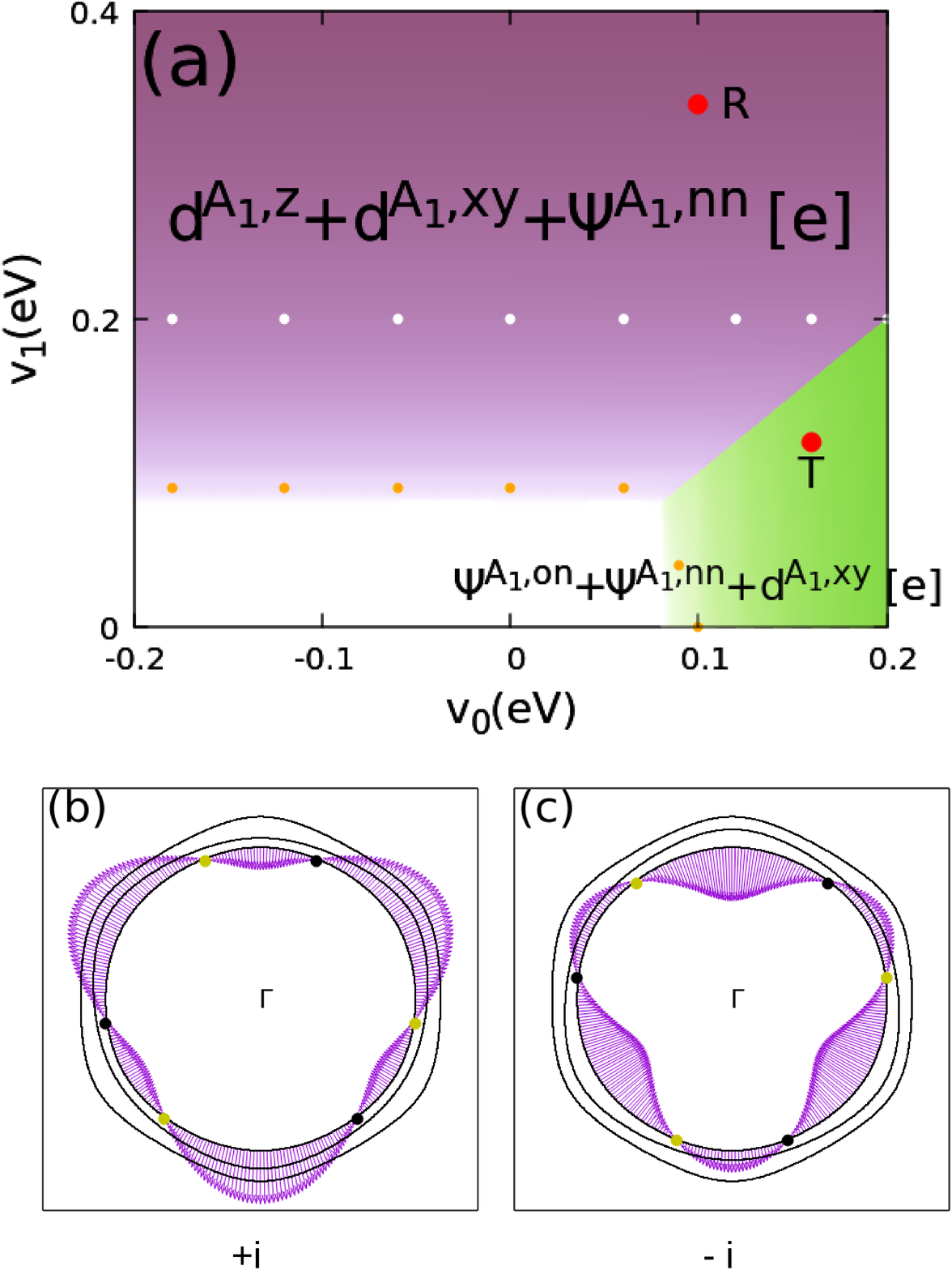}	
		\vspace{-0.2cm}
		\caption{\label{fig04}(color online).(a) The pairing phase diagram with strong Rashba SOC $\alpha_R=50$ meV. The inset is the corresponding band structure. The FS around $\Gamma$ of (b)sector $+i$ and (c)sector $-i$ for the representative new mixing state denoted by `R' in (a). The arrows represent the projected gap functions $\widetilde{\Delta}_{n,n}(\bm{k})$ on the innermost FS, with their lengths(angles) denoting the absolute values(phases) of $\widetilde{\Delta}_{n,n}(\bm{k})$. The black(yellow) dots represent the nodes with WN +1(-1). }
	\end{figure}

	Taking the empirical temperature-dependent upper critical magnetic field of thin layer 2$H$-TaS$_2$, which can not be explained by a pure singlet or triplet pairing, into account\cite{Barrera2018}, the nodal f+s-wave pairing state mentioned above can be a promising candidate. This mixing state has been proposed in the gated superconducting MoS$_2$\cite{Yuan2014,Hsu2017}. It is also consistent with the STM experiment on 2$H$-TaS$_2$, where a ZBCP in the superconducting TaS$_2$ detached flakes was observed\cite{Galvis2014}.

	In the above discussion, the Rashba SOC $\alpha_R$ has been neglected for simplicity. If a relatively small $\alpha_R$ is taken into account, the above main results are qualitatively unchanged. However, if $\alpha_R$ is assumed to be sufficiently large, a $\bm{d}^{A_{1},xy}$ pairing component is expected to be induced, since the Rashba SOC favors the triplet pairing with $\bm{d}^\Gamma \perp \bm{z}$. A detailed calculation confirms this point and gives the pairing phase diagram as shown in Fig. \ref{fig04}(a) for $\alpha_R$ up to $50$ meV, which is so large that it is competing with the intrinsic SOC and hence strongly suppressing Ising pairing. Here $S_z$ conversation is violated, so neither $H_{+i}(\bm{k})$ nor $H_{-i}(\bm{k})$ could be block-diagonalized as before. This system with strong Rashba SOC is found to be still gapless but with reduced number of nodes: All the FS sheets are fully gapped except the two innermost ones around $\Gamma$, each of which has 6 nodes, as depicted in Fig. \ref{fig04}(b)-(c), respectively.

	\begin{figure*}
		\begin{center}
			\includegraphics[width=15cm]{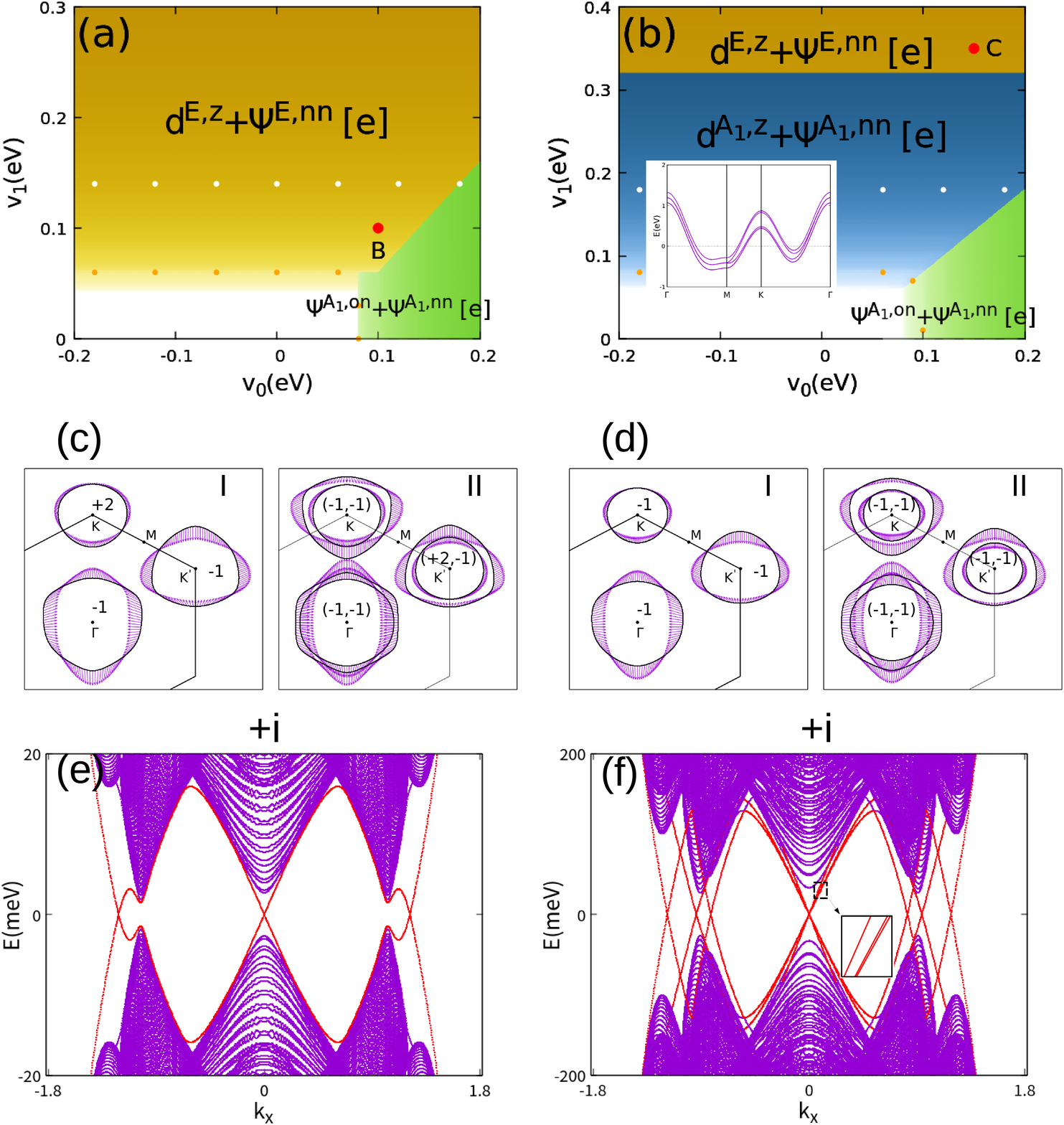}				
		\end{center}
		\vspace{-0.2cm}
		\caption{\label{fig05}(color online).(a)(b)$v_0$ versus $v_1$ pairing phase diagram for the hole-doped and uniaxially compressed trilayer 2$H$-TaS$_2$, respectively. In (a), the chemical potential $\mu$ is set to be -100 meV, while in (b) $t_\perp$ is set to be -90 meV with the carrier density being unchanged. The inset of (b) gives the band structure under uniaxial pressure along z axis. (c)(d)The FS of sector $+i$ for a representative $\bm{d}^{E_1,z}+\Psi^{E_1,nn}$ state denoted by `B' in (a) and by `C' in (b). The number on each FS denotes the phase WN of $\widetilde{\Delta}_{n,n}$. The expansion coefficients for the gap function takes values in Table \ref{tab4} and gap value is chosen to be $\Delta \approx$ 1 K for `B' and 200 K for `C'. (e)(f) The corresponding edge band for sector $+i$ with the open-boundary conditions along y axis, where for better view, the gap value in (e) has been enlarged 20 times.}		
	\end{figure*}

	\subsection{Doping and pressure effects of trilayer 2$H$-TaS$_2$}

	The superconducting behavior can be significantly tuned by doping. Experimentally, the bulk TMDs can be chemically doped with Na and Cu or electron doped by the substrate\cite{Fang2005,Wagner2008,Albertini2017,Lian2017}, while the thin films on a substrate can also be effectively doped by a gate voltage\cite{Lu2015,Xi2016,Lu2018}. Here in the trilayer 2$H$-TaS$_2$, we assume a rigid band and consider the effect of p-type doping by fixing the chemical potential $\mu=-100$ meV, which is near the Van Hove singularity at $M$. Fig. \ref{fig05}(a) gives the paring phase diagram, where a new even-mirror-parity mixing phase $\Psi^{E,nn}+ \bm{d}^{E,z}$ of the IR $E$ appears. Because the IR $E$ is 2D, in this mixing state any combination between $\Psi^{E,nn}_1$ and $\Psi^{E,nn}_2$ (or $\bm{d}^{E,z}_1$ and $\bm{d}^{E,z}_2$) is allowed and shares identical T$_{c}$ determined by Eq.(\ref{eqn15}). In order to determine which combination is the most energetically favorable, one can make an energy minimization, which leads to a TRS breaking mixing state between d+id(d-id)- and p-ip(p+ip)-wave pairings. The detailed expansion coefficients for a representative state for this phase is given in Table \ref{tab4}. This mixing Ising pairing phase is fully gapped. As an example, Fig. \ref{fig05}(c) show the projected gap functions on the FS.  A mirror Chern number(MCN ) $C_{\pm i}$ can be defined for each sector $H_{\pm i}(\bm{k})$. More conveniently, MCN $C^{\Rmnum{1}(\Rmnum{2})}_{\pm i}$ is also meaningful for each subsector $H_{\pm i}^{\Rmnum{1}(\Rmnum{2})}(\bm{k})$, as total spin $S_{z}$ is still conserved in the superconducting state. In the weak-coupling limit, each MCN can actually be viewed as the sum over the phase WNs of the projected order parameter $\widetilde{\Delta}_{n,n}(\bm{k})$ on each FS, as depicted in Fig. \ref{fig05}(c). Here we have a new WN defined on the n-th Fermi pocket $\gamma$:
	
		\begin{eqnarray}
		\nu_n=\frac{1}{2 \pi i}\oint_\gamma dk \partial_{k} ln\widetilde{\Delta}_{n,n}(\bm{k})
		,\end{eqnarray}
		with the circuit integration along the FS.
	Therefore, the total Chern number of the trilayer 2$H$-TaS$_2$ is -6, namely $(C_{+i},C_{-i})=(-3,-3)$, which is consistent with the corresponding chiral edge states of $H_{+i_{}}$ shown in Fig. \ref{fig05}(e).
	
	\renewcommand\arraystretch{1.5}
	\begin{table*}[]
		\caption{The detailed expansion coefficients for the representative mixing pairing states of the 2D IR $E$ corresponding to point `B' and `C' in Fig. \ref{fig05}(a) and (b), respectively. }
		\begin{tabular}{|p{4cm}<{\centering}|p{4cm}<{\centering}|p{4cm}<{\centering}|p{4cm}<{\centering}|c|}
			\hline
			\hline
			$b^{E,nn}_{1}$; $\bm{\chi}^{E,nn}_{1}$ & $b^{E,nn}_{2};\bm{\chi}^{E,nn}_{2}$ & $b^{E,z}_{1}$; $\bm{\chi}^{E,z}_{1}$ & $b^{E,z}_{2};\bm{\chi}^{E,z}_{2}$&State \\
			\hline
			0.285; (-0.637, 0.434,-0.637)&0;$\backslash$&0.958; (0.571, 0.589, 0.571)&0;$\backslash$&$B$ \\
			0.182; (-0.665, 0.341,-0.665)&  0;$\backslash$&0.983; (0.564, 0.602, 0.564)&0;$\backslash$&$C$		\\
			\hline	
		\end{tabular}
		\label{tab4}
	\end{table*}
	\renewcommand\arraystretch{1}	
	On the other hand, since the couplings between TaS$_2$ layers are weak van der Waals forces, an uniaxial pressure can also be applied to tune the features of the material. The pressure dependences of T$_c$ in 2$H$-TMDs have been measured recently\cite{Suderow2005,Tissen2013,Freitas2016,Lian2017,Grasset2018}. Here we assume that the only effect of the uniaxial pressure along z axis is the enhancement of interlayer coupling $t_\perp$. We set $t_\perp=-90$ meV here and get the pairing diagram as shown in Fig. \ref{fig05}(b). There are three mixing phases. Except the trivial one, both the gapless $\Psi^{A_{1},nn}+\bm{d}^{A_1,z}$ and fully gapped $\Psi^{E,nn}_{1}+\bm{d}^{E,z}_{1}$ phases appear. Similar to the doping case, the  $\Psi^{E,nn}_1+ \bm{d}^{E,z}_1$ state under pressure is gapful and TRS breaking, but takes different Chern number, indicating it is topologically different from that in Fig. \ref{fig05}(a). In detail, the MCN of the two subsectors here are $C^{\Rmnum{1}}_{+i}=-3$ and $C^{\Rmnum{2}}_{+i}=-6$(see Fig. \ref{fig05}(d)). The total Chern number is thus -18, as $C_{+i}=C_{-i}=-9$, consistent with the corresponding currents-carrying chiral edge states of $H_{+i}(\bm{k})$ shown in Fig. \ref{fig05}(f). This mixing pairing phase has a rather high T$_{c}$(about 200K) and is expected to be possibly realized by high-pressure experiments on 2H-TaS$_2$ thin flakes.
	
	\section{Conclusion}\label{sec5}
		In summary, based on the linearized gap equation and symmetry analysis, we have obtained the pairing phase diagram of the superconducting monolayer and trilayer TaS$_2$, and found a nodal  f+s-wave state. We suggest that this nodal pairing could be responsible for the anomalous tunneling conductance observed in STM experiments. The nodal structure is so robust that even a strong mirror-symmetric Rashba SOC(up to 50 meV) cannot fully gap the system. Besides, both p-type doping and uniaxial pressure along z axis could induce a TRS breaking mixing state between d+id- and p-ip-wave pairings, which has a large Chern number. Our result indicates that the superconducting trilayer 2$H$-TaS$_2$ could be a promising candidate for realization of topological superconductors. This study will be helpful to understand the unconventional superconductivity in the thin layer 2$H$-TaS$_2$ and other 2D TMDs. However, final determination of the pairing symmetry of this kind of 2D Ising superconductors requires more theoretical and experimental efforts.

	\section{Acknowledgments}
	We thank Jinpeng Xiao, Feng Xiong, Qinli Zhu and Yao Zhou for useful discussions.
	This work is supported by NSFC Project NO.111774126 and 973 Projects No.2015CB921202.
	\\
	\section*{Appendix}
	\renewcommand{\theequation}{A.\arabic{equation}}
	\setcounter{equation}{0}
	In this appendix we first prove the orthogonal relations Eq.(\ref{eqn6}) and then show in detail how to solve the eiqenequation (\ref{eqn15}) for a definite IR $\Gamma$, obtain its T$_c$ and pairing gap function $\Delta(\bm{k}) $ for a multilayer superconductor.
	
	To prove Eq.(\ref{eqn6}), we define $f_{ij}(\Gamma,\alpha,\Gamma^\prime,\beta)$ as follows,
	\begin{widetext}
		\begin{eqnarray}
		f_{ij}(\Gamma,\alpha,\Gamma^\prime, \beta)&\equiv&\sum_{\bm{k}}Tr\{\Delta^{\Gamma,\alpha}_{i}(\bm{k})\Delta^{\dagger \Gamma^\prime,\beta}_{j}(\bm{k})\} \nonumber\\
		& =&\sum_{\bm{k}}Tr\{D^\dagger(R)\Delta^{\Gamma,\alpha}_{i}(\bm{k})D^*(R)D^\tau(R)\Delta^{\dagger \Gamma^\prime,\beta}_{j}(\bm{k})D(R)\}\nonumber\\
		&=&\frac{1}{g}\sum_{R,m,n,\bm{k}}D^{\Gamma}_{im}(R)D^{*\Gamma^\prime}_{jn}(R)Tr\{\Delta^{\Gamma,\alpha}_{m}(R^{-1}\bm{k})\Delta^{\dagger \Gamma^\prime,\beta}_{n}(R^{-1}\bm{k})\} \nonumber\\
		& =&\sum_{m,n}\frac{1}{d^\Gamma}\delta_{ij}\delta_{mn}\delta_{\Gamma\Gamma^\prime}f_{m,n}(\Gamma,\alpha,\Gamma^\prime, \beta)\nonumber\\
		& =&\delta_{ij}\delta_{\Gamma\Gamma^\prime}\frac{1}{d^\Gamma} \sum_mf_{m,m}(\Gamma,\alpha,\Gamma, \beta),
		\end{eqnarray}
	\end{widetext}
	where $g$ is the order of the symmetry group $G$ of the system, and $R$ a group element with $D(R)$ its $2\times2$ spin-rotation representation. The orthogonal relation between different IRs has been used here in the derivation. Remarkably, $f_{ii}(\Gamma,\alpha,\Gamma,\beta)$ is independent of $i$. Moreover, for $\alpha\neq\beta$ in the same $\Gamma$, $f_{ii}(\Gamma,\alpha,\Gamma,\beta)$ is generically zero, so one has:
	\begin{eqnarray}
	f_{ij}(\Gamma,\alpha,\Gamma^\prime, \beta)=\delta_{\Gamma\Gamma^\prime}\delta_{\alpha\beta}\delta_{ij}f(\Gamma,\alpha),
	\end{eqnarray}
	where the positive number $f(\Gamma,\alpha)\equiv f_{ii}(\Gamma,\alpha,\Gamma,\alpha)$ can be renormalized to be $1$ if $\Delta^{\Gamma,\alpha}_{i}(\bm{k})$ has been properly normalized. Thus we come to the orthogonal relations of Eq.(\ref{eqn6}). These orthogonal relations can also be rewritten as,
	\begin{eqnarray}
	\frac{2}{N}\sum_{\bm{k}}\Psi^{\Gamma,\alpha}_{i}\Psi^{*\Gamma^\prime,\beta}_{j}=\delta_{\Gamma\Gamma^\prime}\delta_{\alpha\beta}\delta_{ij}\\
	\frac{2}{N}\sum_{\bm{k}}\bm{d}^{\Gamma,\alpha}_{i}\cdot \bm{d}^{*\Gamma^\prime,\beta}_{j}=\delta_{\Gamma\Gamma^\prime}\delta_{\alpha\beta}\delta_{ij}
	\end{eqnarray}
	which can be easily confirmed by checking Table \ref{tab1} for crystal C$_{3v}$.
	
	Now by making use of the above orthogonal relations we try to solve Eq.(\ref{eqn15}) for a multi-layer system. Multiplying the two sides of the equation by $[\Delta^{* \Gamma,\alpha}_{i}(\bm{k})]_{s_1,s_2}$, then taking trace over spin indices and making a sum over $\bm{k}$, one has:
	\begin{widetext}
		\begin{eqnarray}
		\frac{1}{N}\sum\limits_{\bm{k}}Tr\{\Delta^{\dagger \Gamma,\alpha}_{i}(\bm{k})\Delta_l(\bm{k})\}&=& \frac{-T_c}{N^{2}} \sum\limits_{\omega_n,\bm{k}s_1s_2} [\Delta^{*\Gamma,\alpha}_{i}(\bm{k})]_{s_1s_2}\sum\limits_{\bm{k}^\prime s_3s_4}V_{s_1s_2s_3s_4}(\bm{k},\bm{k}^\prime)[G(k)\Delta(\bm{k}^\prime)G^{\tau}(-k^\prime)]_{\substack{l,l\\s_3s_4}} \nonumber\\
		&=&\frac{-T_c}{N^2} \sum\limits_{\omega_n,\bm{k}s_1s_2}[\Delta^{*\Gamma,\alpha}_{i}(\bm{k})]_{s_1s_2}\sum \limits_{\substack{\bm{k}^\prime s_3s_4 \\\Gamma^\prime,\beta,j}} v^{\Gamma^\prime,\beta} [\Delta^{\Gamma^\prime,\beta}_{j}(\bm{k})]_{s_1s_2}[\Delta^{*\Gamma^\prime,\beta}_{j}(\bm{k}^\prime)]_{s_3s_4} \nonumber\\
		&&\times [G(k)\Delta(\bm{k}^\prime)G^{\tau}(-k^\prime)]_{\substack{l,l\\s_3s_4}} \nonumber\\
		&=&\frac{-T_c}{N^2} \sum \limits_{\omega_n,\bm{k}^\prime,\Gamma^\prime,\beta} v^{\Gamma^\prime,\beta} Tr\{\Delta^{\dagger\Gamma^\prime,\beta}_{i}(\bm{k}^\prime)[G(k)\Delta(\bm{k}^\prime)G^{\tau}(-k^\prime)]_{\substack{l,l}}\}\sum\limits_{\bm{k}}Tr\{\Delta^{\dagger\Gamma,\alpha}_{i}(\bm{k})\Delta^{\Gamma^\prime,\beta}_{i}(\bm{k})\}  \nonumber\\
		&=&\frac{-T_c}{N}  \sum\limits_{\omega_n,\bm{k}^\prime }v^{\Gamma,\alpha}Tr\{ [G(k)\Delta(\bm{k}^\prime)G^{\tau}(-k^\prime)]_{l,l}\Delta^{\dagger \Gamma,\alpha}_{i}(\bm{k}^\prime)\} \nonumber\\
		&=&\frac{-T_c}{N} \sum\limits_{\omega_n,\bm{k} }v^{\Gamma,\alpha}Tr\{ [G(k)\Delta(\bm{k})G^{\tau}(-k)]_{l,l}\Delta^{\dagger \Gamma,\alpha}_{i}(\bm{k})\},
		\end{eqnarray}
	\end{widetext}
	where N is the total number of unit cells of the system. Substitute the expansion of $\Delta(\bm{k})$ Eq.(\ref{eqn9}) into the above equation, one has
	a simplified version of the eigenequation,
	\begin{eqnarray}
	a^{\Gamma,\alpha}_{l,i}=\sum_{l^{\prime}\beta}{Q^{\Gamma,\alpha \beta}_{l,l^{\prime}}a^{\Gamma,\beta}_{l^{\prime},i}}
	,\end{eqnarray}
	where $a^{\Gamma,\alpha}_{l,i}=b^{\Gamma,\alpha} \chi^{\Gamma,\alpha}_{l,i}$. $Q^{\Gamma,\alpha \beta}_{l, l^{\prime}}$ is independent of i and reads,
	
	\begin{widetext}
		\begin{eqnarray}
		Q^{\Gamma,\alpha \beta}_{l, l^{\prime}}&=&\frac{-T_c v^{\Gamma,\alpha}}{N} \sum\limits_{\beta,\bm{k},\omega_n} Tr\{G_{l l^{\prime}}(k) \Delta^{\Gamma,\beta}_{i}(\bm{k}) G_{l l^{\prime}}^{\tau}(-k)\Delta^{\dagger \Gamma,\alpha}_{i}(\bm{k})\} \nonumber\\
		&=&\frac{-T_c v^{\Gamma,\alpha}}{N} \sum_{\bm{k}\sigma \sigma ^{\prime}} \sum_{l_1 l_2} [\widetilde{\Delta}_{l^{\prime},i}^{\Gamma,\beta}(\bm{k})]_{ \substack{l_1 l_2 \\ \sigma \sigma ^{\prime}}}[\widetilde{\Delta}_{l,i}^{\Gamma,\alpha }(\bm{k})]^*_{ \substack{l_1l_2 \\ \sigma \sigma ^{\prime}}}\frac{\tanh[\beta_{c}\xi_{l_1,\sigma} (\bm{k})/2]+\tanh[\beta_{c}\xi_{l_2,\sigma ^{\prime}} (-\bm{k})/2]}{2 [\xi_{l_1,\sigma} (\bm{k})+\xi_{l_2,\sigma ^{\prime}} (-\bm{k})]},
		\end{eqnarray}
	\end{widetext}
	where $l$, $l^{\prime}, l_1$ and $l_2$ are the layer indices, with $l,l^\prime,l_1,l_2=1,2,\cdots,n$, $\sigma,\sigma^\prime$ the spin indices, $G_{l l^{\prime}}(k)$ is the $2\times2$ Matsubara Green's function for the normal state, which takes the form,
	\begin{eqnarray}
	G_{l l^{\prime}}(k)=\sum_{l_1}U_{l l_1}(\bm{k})[i \omega _n -D_{l_1}(\bm{k})]^{-1}[U_{l^{\prime}l_1}(\bm{k})]^{\dagger} \nonumber \\
	.\end{eqnarray}
	Here $U_{ll'}(\bm{k})$ is the $2\times2$ block of the unitary matrix $U(\bm{k})$, which diagonalize $H_0(\bm{k})$: $U^{\dagger}(\bm{k})H_0(\bm{k})U(\bm{k})=D(\bm{k})$. The $2n\times2n$ diagonal matrix $D(\bm{k})$ has the eigenenergies $\xi_{j}(\bm{k})(j=1,2,...,2n)$ of $H_0(\bm{k})$ as its diagonal entries. The $2\times2$ matrix $D_{l}(\bm{k})$ is given by $D_{l}(\bm{k})=\left(
	\begin{array}{cc}
	\xi_{2l-1}(\bm{k}) & 0 \\
	0 & \xi_{2l}(\bm{k})\\
	\end{array}
	\right)$
	with $\xi_{l,\uparrow}(\bm{k})\equiv\xi_{2l-1}(\bm{k})$, $\xi_{l,\downarrow}(\bm{k})\equiv\xi_{2l}(\bm{k})$.  The effective gap matrix $[\widetilde{\Delta}_{l^{\prime},i}^{\Gamma,\alpha}(\bm{k})]_{l_1 l_2}$ is defined as,
	\begin{eqnarray}
	[\widetilde{\Delta}_{l^{\prime},i}^{\Gamma,\alpha}(\bm{k})]_{l_1 l_2}=U_{l^{\prime}l_1}^{\dagger}(\bm{k}) \Delta^{\Gamma,\alpha}_{i}(\bm{k}) U_{l^{\prime}l_2}^{*}(-\bm{k})
	.\end{eqnarray}
	
	For a definite IR $\Gamma$, if the number of different $\alpha$ is $m$, then $Q$ is an $nmd^{\Gamma}\times nmd^{\Gamma}$ matrix. Since $Q$ is independent of index $i$, it is actually a block-diagonalized one, with each block identical to each other. Therefore we only have to solve the eigenequation corresponding to the reduced $nm\times nm$ $Q$ matrix, and then determine its T$_c$ and eigenstate(the expansion coefficients of the gap function).	
	
	Now we demonstrate how to use Mirror symmetry to further reduce the $Q$ matrix. We   	
	take n=2N$_0$(N$_0$ is an integer) as an example. Only $a^{\Gamma,\alpha}_{1},\cdots ,a^{\Gamma,\alpha}_{N_0}$ are independent because the Mirror symmetry ensures that
	\begin{eqnarray}\label{a10}
	a^{\Gamma,\alpha}_{n+1-l}=\eta_M \eta^{\Gamma,\alpha}_{}a^{\Gamma,\alpha}_{l}
	,\end{eqnarray}
	where $\eta_M=\pm 1$ denotes the mirror parity, while $\eta^{\Gamma,\alpha}_{}=+1(-1)$ for $\psi_i^{\Gamma,\alpha}$ or $\bm{d}_i^{\Gamma,\alpha}||\bm{z}$($\bm{d}_i^{\Gamma,\alpha}\perp \bm{z}$).
	Thus we have:
	\begin{eqnarray} \label{a32}
	a^{\Gamma,\alpha}_{l} &=& \sum\limits_{l^{\prime}=1}^{N_0} \sum\limits_{\beta}^{m}{(Q^{\Gamma ,\alpha \beta}_{ l,l^\prime}+\eta_M \eta^{\Gamma,\alpha}_{}Q^{\Gamma,\alpha \beta}_{ l, n+1-l^\prime})a^{\Gamma,\beta}_{l^{\prime}}} \nonumber\\
	&=&\sum_{l^{\prime}=1}^{N_0} \sum_{\beta}^{m}{\widetilde{Q}^{\Gamma ,\alpha \beta}_{ l,l^\prime}a^{\Gamma,\beta}_{l^{\prime}}}\\
	\widetilde{Q}^{\Gamma,\alpha \beta}_{l,l^{\prime}}
	&=&Q^{\Gamma,\alpha \beta}_{ l,l^\prime}+\eta_M \eta^{\Gamma,\beta}Q^{\Gamma,\alpha  \beta}_{l,n+1-l^{\prime}}
	.\end{eqnarray}	
	
	Thus the reduced $\widetilde{Q}$ is an $N_0m \times N_0m$ matrix.
	
	\bibliography{ref}

\begin{thebibliography}{57}%
\makeatletter
\providecommand \@ifxundefined [1]{%
 \@ifx{#1\undefined}
}%
\providecommand \@ifnum [1]{%
 \ifnum #1\expandafter \@firstoftwo
 \else \expandafter \@secondoftwo
 \fi
}%
\providecommand \@ifx [1]{%
 \ifx #1\expandafter \@firstoftwo
 \else \expandafter \@secondoftwo
 \fi
}%
\providecommand \natexlab [1]{#1}%
\providecommand \enquote  [1]{``#1''}%
\providecommand \bibnamefont  [1]{#1}%
\providecommand \bibfnamefont [1]{#1}%
\providecommand \citenamefont [1]{#1}%
\providecommand \href@noop [0]{\@secondoftwo}%
\providecommand \href [0]{\begingroup \@sanitize@url \@href}%
\providecommand \@href[1]{\@@startlink{#1}\@@href}%
\providecommand \@@href[1]{\endgroup#1\@@endlink}%
\providecommand \@sanitize@url [0]{\catcode `\\12\catcode `\$12\catcode
  `\&12\catcode `\#12\catcode `\^12\catcode `\_12\catcode `\%12\relax}%
\providecommand \@@startlink[1]{}%
\providecommand \@@endlink[0]{}%
\providecommand \url  [0]{\begingroup\@sanitize@url \@url }%
\providecommand \@url [1]{\endgroup\@href {#1}{\urlprefix }}%
\providecommand \urlprefix  [0]{URL }%
\providecommand \Eprint [0]{\href }%
\providecommand \doibase [0]{http://dx.doi.org/}%
\providecommand \selectlanguage [0]{\@gobble}%
\providecommand \bibinfo  [0]{\@secondoftwo}%
\providecommand \bibfield  [0]{\@secondoftwo}%
\providecommand \translation [1]{[#1]}%
\providecommand \BibitemOpen [0]{}%
\providecommand \bibitemStop [0]{}%
\providecommand \bibitemNoStop [0]{.\EOS\space}%
\providecommand \EOS [0]{\spacefactor3000\relax}%
\providecommand \BibitemShut  [1]{\csname bibitem#1\endcsname}%
\let\auto@bib@innerbib\@empty
\bibitem [{\citenamefont {Clayman}\ and\ \citenamefont
  {Frindt}(1971)}]{Clayman1971}%
  \BibitemOpen
  \bibfield  {author} {\bibinfo {author} {\bibfnamefont {B.}~\bibnamefont
  {Clayman}}\ and\ \bibinfo {author} {\bibfnamefont {R.}~\bibnamefont
  {Frindt}},\ }\href {\doibase 10.1016/0038-1098(71)90574-6} {\bibfield
  {journal} {\bibinfo  {journal} {Solid State Commun.}\ }\textbf {\bibinfo
  {volume} {9}},\ \bibinfo {pages} {1881} (\bibinfo {year} {1971})}\BibitemShut
  {NoStop}%
\bibitem [{\citenamefont {Hess}\ \emph {et~al.}(1990)\citenamefont {Hess},
  \citenamefont {Robinson},\ and\ \citenamefont {Waszczak}}]{Hess1990}%
  \BibitemOpen
  \bibfield  {author} {\bibinfo {author} {\bibfnamefont {H.~F.}\ \bibnamefont
  {Hess}}, \bibinfo {author} {\bibfnamefont {R.~B.}\ \bibnamefont {Robinson}},
  \ and\ \bibinfo {author} {\bibfnamefont {J.~V.}\ \bibnamefont {Waszczak}},\
  }\href {\doibase 10.1103/PhysRevLett.64.2711} {\bibfield  {journal} {\bibinfo
   {journal} {Phys. Rev. Lett.}\ }\textbf {\bibinfo {volume} {64}},\ \bibinfo
  {pages} {2711} (\bibinfo {year} {1990})}\BibitemShut {NoStop}%
\bibitem [{\citenamefont {Corcoran}\ \emph {et~al.}(1994)\citenamefont
  {Corcoran}, \citenamefont {Meeson}, \citenamefont {Onuki}, \citenamefont
  {Probst}, \citenamefont {Springford}, \citenamefont {Takita}, \citenamefont
  {Harima}, \citenamefont {Guo},\ and\ \citenamefont {Gyorffy}}]{Corcoran1994}%
  \BibitemOpen
  \bibfield  {author} {\bibinfo {author} {\bibfnamefont {R.}~\bibnamefont
  {Corcoran}}, \bibinfo {author} {\bibfnamefont {P.}~\bibnamefont {Meeson}},
  \bibinfo {author} {\bibfnamefont {Y.}~\bibnamefont {Onuki}}, \bibinfo
  {author} {\bibfnamefont {P.-A.}\ \bibnamefont {Probst}}, \bibinfo {author}
  {\bibfnamefont {M.}~\bibnamefont {Springford}}, \bibinfo {author}
  {\bibfnamefont {K.}~\bibnamefont {Takita}}, \bibinfo {author} {\bibfnamefont
  {H.}~\bibnamefont {Harima}}, \bibinfo {author} {\bibfnamefont
  {G.}~\bibnamefont {Guo}}, \ and\ \bibinfo {author} {\bibfnamefont
  {B.}~\bibnamefont {Gyorffy}},\ }\href {\doibase 10.1088/0953-8984/6/24/010}
  {\bibfield  {journal} {\bibinfo  {journal} {J. Phys. Condens. Matter}\
  }\textbf {\bibinfo {volume} {6}},\ \bibinfo {pages} {4479} (\bibinfo {year}
  {1994})}\BibitemShut {NoStop}%
\bibitem [{\citenamefont {Boaknin}\ \emph {et~al.}(2003)\citenamefont
  {Boaknin}, \citenamefont {Tanatar}, \citenamefont {Paglione}, \citenamefont
  {Hawthorn}, \citenamefont {Ronning}, \citenamefont {Hill}, \citenamefont
  {Sutherland}, \citenamefont {Taillefer}, \citenamefont {Sonier},
  \citenamefont {Hayden},\ and\ \citenamefont {Brill}}]{Boaknin2003}%
  \BibitemOpen
  \bibfield  {author} {\bibinfo {author} {\bibfnamefont {E.}~\bibnamefont
  {Boaknin}}, \bibinfo {author} {\bibfnamefont {M.~A.}\ \bibnamefont
  {Tanatar}}, \bibinfo {author} {\bibfnamefont {J.}~\bibnamefont {Paglione}},
  \bibinfo {author} {\bibfnamefont {D.}~\bibnamefont {Hawthorn}}, \bibinfo
  {author} {\bibfnamefont {F.}~\bibnamefont {Ronning}}, \bibinfo {author}
  {\bibfnamefont {R.~W.}\ \bibnamefont {Hill}}, \bibinfo {author}
  {\bibfnamefont {M.}~\bibnamefont {Sutherland}}, \bibinfo {author}
  {\bibfnamefont {L.}~\bibnamefont {Taillefer}}, \bibinfo {author}
  {\bibfnamefont {J.}~\bibnamefont {Sonier}}, \bibinfo {author} {\bibfnamefont
  {S.~M.}\ \bibnamefont {Hayden}}, \ and\ \bibinfo {author} {\bibfnamefont
  {J.~W.}\ \bibnamefont {Brill}},\ }\href {\doibase
  10.1103/PhysRevLett.90.117003} {\bibfield  {journal} {\bibinfo  {journal}
  {Phys. Rev. Lett.}\ }\textbf {\bibinfo {volume} {90}},\ \bibinfo {pages}
  {117003} (\bibinfo {year} {2003})}\BibitemShut {NoStop}%
\bibitem [{\citenamefont {Huang}\ \emph {et~al.}(2007)\citenamefont {Huang},
  \citenamefont {Lin}, \citenamefont {Chang}, \citenamefont {Sun},
  \citenamefont {Shen}, \citenamefont {Chou}, \citenamefont {Berger},
  \citenamefont {Lee},\ and\ \citenamefont {Yang}}]{Huang2007}%
  \BibitemOpen
  \bibfield  {author} {\bibinfo {author} {\bibfnamefont {C.~L.}\ \bibnamefont
  {Huang}}, \bibinfo {author} {\bibfnamefont {J.-Y.}\ \bibnamefont {Lin}},
  \bibinfo {author} {\bibfnamefont {Y.~T.}\ \bibnamefont {Chang}}, \bibinfo
  {author} {\bibfnamefont {C.~P.}\ \bibnamefont {Sun}}, \bibinfo {author}
  {\bibfnamefont {H.~Y.}\ \bibnamefont {Shen}}, \bibinfo {author}
  {\bibfnamefont {C.~C.}\ \bibnamefont {Chou}}, \bibinfo {author}
  {\bibfnamefont {H.}~\bibnamefont {Berger}}, \bibinfo {author} {\bibfnamefont
  {T.~K.}\ \bibnamefont {Lee}}, \ and\ \bibinfo {author} {\bibfnamefont
  {H.~D.}\ \bibnamefont {Yang}},\ }\href {\doibase 10.1103/PhysRevB.76.212504}
  {\bibfield  {journal} {\bibinfo  {journal} {Phys. Rev. B}\ }\textbf {\bibinfo
  {volume} {76}},\ \bibinfo {pages} {212504} (\bibinfo {year}
  {2007})}\BibitemShut {NoStop}%
\bibitem [{\citenamefont {Berthier}\ \emph {et~al.}(1976)\citenamefont
  {Berthier}, \citenamefont {Molini{\'e}},\ and\ \citenamefont
  {J{\'e}rome}}]{Berthier1976}%
  \BibitemOpen
  \bibfield  {author} {\bibinfo {author} {\bibfnamefont {C.}~\bibnamefont
  {Berthier}}, \bibinfo {author} {\bibfnamefont {P.}~\bibnamefont
  {Molini{\'e}}}, \ and\ \bibinfo {author} {\bibfnamefont {D.}~\bibnamefont
  {J{\'e}rome}},\ }\href {\doibase 10.1016/0038-1098(76)90986-8} {\bibfield
  {journal} {\bibinfo  {journal} {Solid State Commun.}\ }\textbf {\bibinfo
  {volume} {18}},\ \bibinfo {pages} {1393} (\bibinfo {year}
  {1976})}\BibitemShut {NoStop}%
\bibitem [{\citenamefont {Mutka}(1983)}]{Mutka1983}%
  \BibitemOpen
  \bibfield  {author} {\bibinfo {author} {\bibfnamefont {H.}~\bibnamefont
  {Mutka}},\ }\href {\doibase 10.1103/PhysRevB.28.2855} {\bibfield  {journal}
  {\bibinfo  {journal} {Phys. Rev. B}\ }\textbf {\bibinfo {volume} {28}},\
  \bibinfo {pages} {2855} (\bibinfo {year} {1983})}\BibitemShut {NoStop}%
\bibitem [{\citenamefont {Castro~Neto}(2001)}]{Neto2001}%
  \BibitemOpen
  \bibfield  {author} {\bibinfo {author} {\bibfnamefont {A.~H.}\ \bibnamefont
  {Castro~Neto}},\ }\href {\doibase 10.1103/PhysRevLett.86.4382} {\bibfield
  {journal} {\bibinfo  {journal} {Phys. Rev. Lett.}\ }\textbf {\bibinfo
  {volume} {86}},\ \bibinfo {pages} {4382} (\bibinfo {year}
  {2001})}\BibitemShut {NoStop}%
\bibitem [{\citenamefont {Yokoya}\ \emph {et~al.}(2001)\citenamefont {Yokoya},
  \citenamefont {Kiss}, \citenamefont {Chainani}, \citenamefont {Shin},
  \citenamefont {Nohara},\ and\ \citenamefont {Takagi}}]{Yokoya2001}%
  \BibitemOpen
  \bibfield  {author} {\bibinfo {author} {\bibfnamefont {T.}~\bibnamefont
  {Yokoya}}, \bibinfo {author} {\bibfnamefont {T.}~\bibnamefont {Kiss}},
  \bibinfo {author} {\bibfnamefont {A.}~\bibnamefont {Chainani}}, \bibinfo
  {author} {\bibfnamefont {S.}~\bibnamefont {Shin}}, \bibinfo {author}
  {\bibfnamefont {M.}~\bibnamefont {Nohara}}, \ and\ \bibinfo {author}
  {\bibfnamefont {H.}~\bibnamefont {Takagi}},\ }\href {\doibase
  10.1126/science.1065068} {\bibfield  {journal} {\bibinfo  {journal}
  {Science}\ }\textbf {\bibinfo {volume} {294}},\ \bibinfo {pages} {2518}
  (\bibinfo {year} {2001})}\BibitemShut {NoStop}%
\bibitem [{\citenamefont {Guillam{\'o}n}\ \emph {et~al.}(2011)\citenamefont
  {Guillam{\'o}n}, \citenamefont {Suderow}, \citenamefont {Rodrigo},
  \citenamefont {Vieira}, \citenamefont {Rodiere}, \citenamefont {Cario},
  \citenamefont {Navarro-Moratalla}, \citenamefont {Mart{\'\i}-Gastaldo},\ and\
  \citenamefont {Coronado}}]{Guillamon2011}%
  \BibitemOpen
  \bibfield  {author} {\bibinfo {author} {\bibfnamefont {I.}~\bibnamefont
  {Guillam{\'o}n}}, \bibinfo {author} {\bibfnamefont {H.}~\bibnamefont
  {Suderow}}, \bibinfo {author} {\bibfnamefont {J.~G.}\ \bibnamefont
  {Rodrigo}}, \bibinfo {author} {\bibfnamefont {S.}~\bibnamefont {Vieira}},
  \bibinfo {author} {\bibfnamefont {P.}~\bibnamefont {Rodiere}}, \bibinfo
  {author} {\bibfnamefont {L.}~\bibnamefont {Cario}}, \bibinfo {author}
  {\bibfnamefont {E.}~\bibnamefont {Navarro-Moratalla}}, \bibinfo {author}
  {\bibfnamefont {C.}~\bibnamefont {Mart{\'\i}-Gastaldo}}, \ and\ \bibinfo
  {author} {\bibfnamefont {E.}~\bibnamefont {Coronado}},\ }\href {\doibase
  10.1088/1367-2630/13/10/103020} {\bibfield  {journal} {\bibinfo  {journal}
  {New J. Phys.}\ }\textbf {\bibinfo {volume} {13}},\ \bibinfo {pages} {103020}
  (\bibinfo {year} {2011})}\BibitemShut {NoStop}%
\bibitem [{\citenamefont {Xi}\ \emph {et~al.}(2015{\natexlab{a}})\citenamefont
  {Xi}, \citenamefont {Zhao}, \citenamefont {Wang}, \citenamefont {Berger},
  \citenamefont {Forr{\'o}}, \citenamefont {Shan},\ and\ \citenamefont
  {Mak}}]{Xi2015N}%
  \BibitemOpen
  \bibfield  {author} {\bibinfo {author} {\bibfnamefont {X.}~\bibnamefont
  {Xi}}, \bibinfo {author} {\bibfnamefont {L.}~\bibnamefont {Zhao}}, \bibinfo
  {author} {\bibfnamefont {Z.}~\bibnamefont {Wang}}, \bibinfo {author}
  {\bibfnamefont {H.}~\bibnamefont {Berger}}, \bibinfo {author} {\bibfnamefont
  {L.}~\bibnamefont {Forr{\'o}}}, \bibinfo {author} {\bibfnamefont
  {J.}~\bibnamefont {Shan}}, \ and\ \bibinfo {author} {\bibfnamefont {K.~F.}\
  \bibnamefont {Mak}},\ }\href {\doibase 10.1038/nnano.2015.143} {\bibfield
  {journal} {\bibinfo  {journal} {Nat. nanotechnol.}\ }\textbf {\bibinfo
  {volume} {10}},\ \bibinfo {pages} {765} (\bibinfo {year}
  {2015}{\natexlab{a}})}\BibitemShut {NoStop}%
\bibitem [{\citenamefont {Xiao}\ \emph {et~al.}(2012)\citenamefont {Xiao},
  \citenamefont {Liu}, \citenamefont {Feng}, \citenamefont {Xu},\ and\
  \citenamefont {Yao}}]{Xiao2012}%
  \BibitemOpen
  \bibfield  {author} {\bibinfo {author} {\bibfnamefont {D.}~\bibnamefont
  {Xiao}}, \bibinfo {author} {\bibfnamefont {G.-B.}\ \bibnamefont {Liu}},
  \bibinfo {author} {\bibfnamefont {W.}~\bibnamefont {Feng}}, \bibinfo {author}
  {\bibfnamefont {X.}~\bibnamefont {Xu}}, \ and\ \bibinfo {author}
  {\bibfnamefont {W.}~\bibnamefont {Yao}},\ }\href {\doibase
  10.1103/PhysRevLett.108.196802} {\bibfield  {journal} {\bibinfo  {journal}
  {Phys. Rev. Lett.}\ }\textbf {\bibinfo {volume} {108}},\ \bibinfo {pages}
  {196802} (\bibinfo {year} {2012})}\BibitemShut {NoStop}%
\bibitem [{\citenamefont {Lu}\ \emph {et~al.}(2013)\citenamefont {Lu},
  \citenamefont {Yao}, \citenamefont {Xiao},\ and\ \citenamefont
  {Shen}}]{Lu2013}%
  \BibitemOpen
  \bibfield  {author} {\bibinfo {author} {\bibfnamefont {H.-Z.}\ \bibnamefont
  {Lu}}, \bibinfo {author} {\bibfnamefont {W.}~\bibnamefont {Yao}}, \bibinfo
  {author} {\bibfnamefont {D.}~\bibnamefont {Xiao}}, \ and\ \bibinfo {author}
  {\bibfnamefont {S.-Q.}\ \bibnamefont {Shen}},\ }\href {\doibase
  10.1103/PhysRevLett.110.016806} {\bibfield  {journal} {\bibinfo  {journal}
  {Phys. Rev. Lett.}\ }\textbf {\bibinfo {volume} {110}},\ \bibinfo {pages}
  {016806} (\bibinfo {year} {2013})}\BibitemShut {NoStop}%
\bibitem [{\citenamefont {Suzuki}\ \emph {et~al.}(2014)\citenamefont {Suzuki},
  \citenamefont {Sakano}, \citenamefont {Zhang}, \citenamefont {Akashi},
  \citenamefont {Morikawa}, \citenamefont {Harasawa}, \citenamefont {Yaji},
  \citenamefont {Kuroda}, \citenamefont {Miyamoto}, \citenamefont {Okuda} \emph
  {et~al.}}]{Suzuki2014}%
  \BibitemOpen
  \bibfield  {author} {\bibinfo {author} {\bibfnamefont {R.}~\bibnamefont
  {Suzuki}}, \bibinfo {author} {\bibfnamefont {M.}~\bibnamefont {Sakano}},
  \bibinfo {author} {\bibfnamefont {Y.}~\bibnamefont {Zhang}}, \bibinfo
  {author} {\bibfnamefont {R.}~\bibnamefont {Akashi}}, \bibinfo {author}
  {\bibfnamefont {D.}~\bibnamefont {Morikawa}}, \bibinfo {author}
  {\bibfnamefont {A.}~\bibnamefont {Harasawa}}, \bibinfo {author}
  {\bibfnamefont {K.}~\bibnamefont {Yaji}}, \bibinfo {author} {\bibfnamefont
  {K.}~\bibnamefont {Kuroda}}, \bibinfo {author} {\bibfnamefont
  {K.}~\bibnamefont {Miyamoto}}, \bibinfo {author} {\bibfnamefont
  {T.}~\bibnamefont {Okuda}},  \emph {et~al.},\ }\href {\doibase
  10.1038/nphys3580} {\bibfield  {journal} {\bibinfo  {journal} {Nat.
  nanotechnol.}\ }\textbf {\bibinfo {volume} {9}},\ \bibinfo {pages} {611}
  (\bibinfo {year} {2014})}\BibitemShut {NoStop}%
\bibitem [{\citenamefont {Bawden}\ \emph {et~al.}(2016)\citenamefont {Bawden},
  \citenamefont {Cooil}, \citenamefont {Mazzola}, \citenamefont {Riley},
  \citenamefont {Collins-McIntyre}, \citenamefont {Sunko}, \citenamefont
  {Hunvik}, \citenamefont {Leandersson}, \citenamefont {Polley}, \citenamefont
  {Balasubramanian} \emph {et~al.}}]{Bawden2016}%
  \BibitemOpen
  \bibfield  {author} {\bibinfo {author} {\bibfnamefont {L.}~\bibnamefont
  {Bawden}}, \bibinfo {author} {\bibfnamefont {S.}~\bibnamefont {Cooil}},
  \bibinfo {author} {\bibfnamefont {F.}~\bibnamefont {Mazzola}}, \bibinfo
  {author} {\bibfnamefont {J.}~\bibnamefont {Riley}}, \bibinfo {author}
  {\bibfnamefont {L.}~\bibnamefont {Collins-McIntyre}}, \bibinfo {author}
  {\bibfnamefont {V.}~\bibnamefont {Sunko}}, \bibinfo {author} {\bibfnamefont
  {K.}~\bibnamefont {Hunvik}}, \bibinfo {author} {\bibfnamefont
  {M.}~\bibnamefont {Leandersson}}, \bibinfo {author} {\bibfnamefont
  {C.}~\bibnamefont {Polley}}, \bibinfo {author} {\bibfnamefont
  {T.}~\bibnamefont {Balasubramanian}},  \emph {et~al.},\ }\href {\doibase
  10.1038/ncomms11711} {\bibfield  {journal} {\bibinfo  {journal} {Nat.
  Commun.}\ }\textbf {\bibinfo {volume} {7}},\ \bibinfo {pages} {11711}
  (\bibinfo {year} {2016})}\BibitemShut {NoStop}%
\bibitem [{\citenamefont {Wu}\ \emph {et~al.}(2016)\citenamefont {Wu},
  \citenamefont {Xu}, \citenamefont {Lu}, \citenamefont {Khamoshi},
  \citenamefont {Liu}, \citenamefont {Han}, \citenamefont {Wu}, \citenamefont
  {Lin}, \citenamefont {Long},\ and\ \citenamefont {He}}]{Wu2016}%
  \BibitemOpen
  \bibfield  {author} {\bibinfo {author} {\bibfnamefont {Z.}~\bibnamefont
  {Wu}}, \bibinfo {author} {\bibfnamefont {S.}~\bibnamefont {Xu}}, \bibinfo
  {author} {\bibfnamefont {H.}~\bibnamefont {Lu}}, \bibinfo {author}
  {\bibfnamefont {A.}~\bibnamefont {Khamoshi}}, \bibinfo {author}
  {\bibfnamefont {G.~B.}\ \bibnamefont {Liu}}, \bibinfo {author} {\bibfnamefont
  {T.}~\bibnamefont {Han}}, \bibinfo {author} {\bibfnamefont {Y.}~\bibnamefont
  {Wu}}, \bibinfo {author} {\bibfnamefont {J.}~\bibnamefont {Lin}}, \bibinfo
  {author} {\bibfnamefont {G.}~\bibnamefont {Long}}, \ and\ \bibinfo {author}
  {\bibfnamefont {Y.}~\bibnamefont {He}},\ }\href {\doibase
  10.1038/ncomms12955} {\bibfield  {journal} {\bibinfo  {journal} {Nat.
  Commun.}\ }\textbf {\bibinfo {volume} {7}},\ \bibinfo {pages} {12955}
  (\bibinfo {year} {2016})}\BibitemShut {NoStop}%
\bibitem [{\citenamefont {Dey}\ \emph {et~al.}(2017)\citenamefont {Dey},
  \citenamefont {Yang}, \citenamefont {Robert}, \citenamefont {Wang},
  \citenamefont {Urbaszek}, \citenamefont {Marie},\ and\ \citenamefont
  {Crooker}}]{Dey2017}%
  \BibitemOpen
  \bibfield  {author} {\bibinfo {author} {\bibfnamefont {P.}~\bibnamefont
  {Dey}}, \bibinfo {author} {\bibfnamefont {L.}~\bibnamefont {Yang}}, \bibinfo
  {author} {\bibfnamefont {C.}~\bibnamefont {Robert}}, \bibinfo {author}
  {\bibfnamefont {G.}~\bibnamefont {Wang}}, \bibinfo {author} {\bibfnamefont
  {B.}~\bibnamefont {Urbaszek}}, \bibinfo {author} {\bibfnamefont
  {X.}~\bibnamefont {Marie}}, \ and\ \bibinfo {author} {\bibfnamefont {S.~A.}\
  \bibnamefont {Crooker}},\ }\href {\doibase 10.1103/PhysRevLett.119.137401}
  {\bibfield  {journal} {\bibinfo  {journal} {Phys. Rev. Lett.}\ }\textbf
  {\bibinfo {volume} {119}},\ \bibinfo {pages} {137401} (\bibinfo {year}
  {2017})}\BibitemShut {NoStop}%
\bibitem [{\citenamefont {Mak}\ \emph {et~al.}(2014)\citenamefont {Mak},
  \citenamefont {McGill}, \citenamefont {Park},\ and\ \citenamefont
  {McEuen}}]{Mak2014}%
  \BibitemOpen
  \bibfield  {author} {\bibinfo {author} {\bibfnamefont {K.~F.}\ \bibnamefont
  {Mak}}, \bibinfo {author} {\bibfnamefont {K.~L.}\ \bibnamefont {McGill}},
  \bibinfo {author} {\bibfnamefont {J.}~\bibnamefont {Park}}, \ and\ \bibinfo
  {author} {\bibfnamefont {P.~L.}\ \bibnamefont {McEuen}},\ }\href {\doibase
  10.1126/science.1250140} {\bibfield  {journal} {\bibinfo  {journal}
  {Science}\ }\textbf {\bibinfo {volume} {344}},\ \bibinfo {pages} {1489}
  (\bibinfo {year} {2014})}\BibitemShut {NoStop}%
\bibitem [{\citenamefont {Lee}\ \emph {et~al.}(2016)\citenamefont {Lee},
  \citenamefont {Mak},\ and\ \citenamefont {Shan}}]{Lee2016}%
  \BibitemOpen
  \bibfield  {author} {\bibinfo {author} {\bibfnamefont {J.}~\bibnamefont
  {Lee}}, \bibinfo {author} {\bibfnamefont {K.~F.}\ \bibnamefont {Mak}}, \ and\
  \bibinfo {author} {\bibfnamefont {J.}~\bibnamefont {Shan}},\ }\href {\doibase
  10.1038/nnano.2015.337} {\bibfield  {journal} {\bibinfo  {journal} {Nat.
  Nanotechnol.}\ }\textbf {\bibinfo {volume} {11}},\ \bibinfo {pages} {421}
  (\bibinfo {year} {2016})}\BibitemShut {NoStop}%
\bibitem [{\citenamefont {Yu}\ and\ \citenamefont {Wu}(2016)}]{Yu2016}%
  \BibitemOpen
  \bibfield  {author} {\bibinfo {author} {\bibfnamefont {T.}~\bibnamefont
  {Yu}}\ and\ \bibinfo {author} {\bibfnamefont {M.~W.}\ \bibnamefont {Wu}},\
  }\href {\doibase 10.1103/PhysRevB.93.045414} {\bibfield  {journal} {\bibinfo
  {journal} {Phys. Rev. B}\ }\textbf {\bibinfo {volume} {93}},\ \bibinfo
  {pages} {045414} (\bibinfo {year} {2016})}\BibitemShut {NoStop}%
\bibitem [{\citenamefont {JM}\ \emph {et~al.}(2015)\citenamefont {JM},
  \citenamefont {O}, \citenamefont {I}, \citenamefont {NF}, \citenamefont {U},
  \citenamefont {KT},\ and\ \citenamefont {JT}}]{Lu2015}%
  \BibitemOpen
  \bibfield  {author} {\bibinfo {author} {\bibfnamefont {L.}~\bibnamefont
  {JM}}, \bibinfo {author} {\bibfnamefont {Z.}~\bibnamefont {O}}, \bibinfo
  {author} {\bibfnamefont {L.}~\bibnamefont {I}}, \bibinfo {author}
  {\bibfnamefont {Y.}~\bibnamefont {NF}}, \bibinfo {author} {\bibfnamefont
  {Z.}~\bibnamefont {U}}, \bibinfo {author} {\bibfnamefont {L.}~\bibnamefont
  {KT}}, \ and\ \bibinfo {author} {\bibfnamefont {Y.}~\bibnamefont {JT}},\
  }\href {\doibase 10.1126/science.aab2277} {\bibfield  {journal} {\bibinfo
  {journal} {Science}\ }\textbf {\bibinfo {volume} {350}},\ \bibinfo {pages}
  {1353} (\bibinfo {year} {2015})}\BibitemShut {NoStop}%
\bibitem [{\citenamefont {Xi}\ \emph {et~al.}(2015{\natexlab{b}})\citenamefont
  {Xi}, \citenamefont {Wang}, \citenamefont {Zhao}, \citenamefont {Park},
  \citenamefont {Tuen~Law}, \citenamefont {Berger}, \citenamefont {Forró},
  \citenamefont {Shan},\ and\ \citenamefont {Mak}}]{Xi2015}%
  \BibitemOpen
  \bibfield  {author} {\bibinfo {author} {\bibfnamefont {X.}~\bibnamefont
  {Xi}}, \bibinfo {author} {\bibfnamefont {Z.}~\bibnamefont {Wang}}, \bibinfo
  {author} {\bibfnamefont {W.}~\bibnamefont {Zhao}}, \bibinfo {author}
  {\bibfnamefont {J.-H.}\ \bibnamefont {Park}}, \bibinfo {author}
  {\bibfnamefont {K.}~\bibnamefont {Tuen~Law}}, \bibinfo {author}
  {\bibfnamefont {H.}~\bibnamefont {Berger}}, \bibinfo {author} {\bibfnamefont
  {L.}~\bibnamefont {Forró}}, \bibinfo {author} {\bibfnamefont
  {J.}~\bibnamefont {Shan}}, \ and\ \bibinfo {author} {\bibfnamefont
  {K.}~\bibnamefont {Mak}},\ }\href {\doibase 10.1038/nphys3538} {\bibfield
  {journal} {\bibinfo  {journal} {Nat. Phys.}\ }\textbf {\bibinfo {volume}
  {12}},\ \bibinfo {pages} {139} (\bibinfo {year}
  {2015}{\natexlab{b}})}\BibitemShut {NoStop}%
\bibitem [{\citenamefont {Saito}\ \emph {et~al.}(2016)\citenamefont {Saito},
  \citenamefont {Nakamura}, \citenamefont {Bahramy}, \citenamefont {Kohama},
  \citenamefont {Ye}, \citenamefont {Kasahara}, \citenamefont {Nakagawa},
  \citenamefont {Onga}, \citenamefont {Tokunaga}, \citenamefont {Nojima} \emph
  {et~al.}}]{Saito2016}%
  \BibitemOpen
  \bibfield  {author} {\bibinfo {author} {\bibfnamefont {Y.}~\bibnamefont
  {Saito}}, \bibinfo {author} {\bibfnamefont {Y.}~\bibnamefont {Nakamura}},
  \bibinfo {author} {\bibfnamefont {M.~S.}\ \bibnamefont {Bahramy}}, \bibinfo
  {author} {\bibfnamefont {Y.}~\bibnamefont {Kohama}}, \bibinfo {author}
  {\bibfnamefont {J.}~\bibnamefont {Ye}}, \bibinfo {author} {\bibfnamefont
  {Y.}~\bibnamefont {Kasahara}}, \bibinfo {author} {\bibfnamefont
  {Y.}~\bibnamefont {Nakagawa}}, \bibinfo {author} {\bibfnamefont
  {M.}~\bibnamefont {Onga}}, \bibinfo {author} {\bibfnamefont {M.}~\bibnamefont
  {Tokunaga}}, \bibinfo {author} {\bibfnamefont {T.}~\bibnamefont {Nojima}},
  \emph {et~al.},\ }\href {\doibase 10.1038/nphys3580} {\bibfield  {journal}
  {\bibinfo  {journal} {Nat. Phys.}\ }\textbf {\bibinfo {volume} {12}},\
  \bibinfo {pages} {144} (\bibinfo {year} {2016})}\BibitemShut {NoStop}%
\bibitem [{\citenamefont {Xing}\ \emph {et~al.}(2017)\citenamefont {Xing},
  \citenamefont {Zhao}, \citenamefont {Shan}, \citenamefont {Zheng},
  \citenamefont {Zhang}, \citenamefont {Fu}, \citenamefont {Liu}, \citenamefont
  {Tian}, \citenamefont {Xi}, \citenamefont {Liu}, \citenamefont {Feng},
  \citenamefont {Lin}, \citenamefont {Ji}, \citenamefont {Chen}, \citenamefont
  {Xue},\ and\ \citenamefont {Wang}}]{Xing2017}%
  \BibitemOpen
  \bibfield  {author} {\bibinfo {author} {\bibfnamefont {Y.}~\bibnamefont
  {Xing}}, \bibinfo {author} {\bibfnamefont {K.}~\bibnamefont {Zhao}}, \bibinfo
  {author} {\bibfnamefont {P.}~\bibnamefont {Shan}}, \bibinfo {author}
  {\bibfnamefont {F.}~\bibnamefont {Zheng}}, \bibinfo {author} {\bibfnamefont
  {Y.}~\bibnamefont {Zhang}}, \bibinfo {author} {\bibfnamefont
  {H.}~\bibnamefont {Fu}}, \bibinfo {author} {\bibfnamefont {Y.}~\bibnamefont
  {Liu}}, \bibinfo {author} {\bibfnamefont {M.}~\bibnamefont {Tian}}, \bibinfo
  {author} {\bibfnamefont {C.}~\bibnamefont {Xi}}, \bibinfo {author}
  {\bibfnamefont {H.}~\bibnamefont {Liu}}, \bibinfo {author} {\bibfnamefont
  {J.}~\bibnamefont {Feng}}, \bibinfo {author} {\bibfnamefont {X.}~\bibnamefont
  {Lin}}, \bibinfo {author} {\bibfnamefont {S.}~\bibnamefont {Ji}}, \bibinfo
  {author} {\bibfnamefont {X.}~\bibnamefont {Chen}}, \bibinfo {author}
  {\bibfnamefont {Q.-K.}\ \bibnamefont {Xue}}, \ and\ \bibinfo {author}
  {\bibfnamefont {J.}~\bibnamefont {Wang}},\ }\href {\doibase
  10.1021/acs.nanolett.7b03026} {\bibfield  {journal} {\bibinfo  {journal}
  {Nano Lett.}\ }\textbf {\bibinfo {volume} {17}},\ \bibinfo {pages} {6802}
  (\bibinfo {year} {2017})}\BibitemShut {NoStop}%
\bibitem [{\citenamefont {Lu}\ \emph {et~al.}(2018)\citenamefont {Lu},
  \citenamefont {Zheliuk}, \citenamefont {Chen}, \citenamefont {Leermakers},
  \citenamefont {Hussey}, \citenamefont {Zeitler},\ and\ \citenamefont
  {Ye}}]{Lu2018}%
  \BibitemOpen
  \bibfield  {author} {\bibinfo {author} {\bibfnamefont {J.}~\bibnamefont
  {Lu}}, \bibinfo {author} {\bibfnamefont {O.}~\bibnamefont {Zheliuk}},
  \bibinfo {author} {\bibfnamefont {Q.}~\bibnamefont {Chen}}, \bibinfo {author}
  {\bibfnamefont {I.}~\bibnamefont {Leermakers}}, \bibinfo {author}
  {\bibfnamefont {N.~E.}\ \bibnamefont {Hussey}}, \bibinfo {author}
  {\bibfnamefont {U.}~\bibnamefont {Zeitler}}, \ and\ \bibinfo {author}
  {\bibfnamefont {J.}~\bibnamefont {Ye}},\ }\href {\doibase
  10.1073/pnas.1716781115} {\bibfield  {journal} {\bibinfo  {journal} {Proc.
  Natl. Acad. Sci. U.S.A.}\ }\textbf {\bibinfo {volume} {115}},\ \bibinfo
  {pages} {3551} (\bibinfo {year} {2018})}\BibitemShut {NoStop}%
\bibitem [{\citenamefont {Sohn}\ \emph {et~al.}(2018)\citenamefont {Sohn},
  \citenamefont {Xi}, \citenamefont {He}, \citenamefont {Jiang}, \citenamefont
  {Wang}, \citenamefont {Kang}, \citenamefont {Park}, \citenamefont {Berger},
  \citenamefont {Forr{\'o}}, \citenamefont {Law} \emph {et~al.}}]{Sohn2018}%
  \BibitemOpen
  \bibfield  {author} {\bibinfo {author} {\bibfnamefont {E.}~\bibnamefont
  {Sohn}}, \bibinfo {author} {\bibfnamefont {X.}~\bibnamefont {Xi}}, \bibinfo
  {author} {\bibfnamefont {W.-Y.}\ \bibnamefont {He}}, \bibinfo {author}
  {\bibfnamefont {S.}~\bibnamefont {Jiang}}, \bibinfo {author} {\bibfnamefont
  {Z.}~\bibnamefont {Wang}}, \bibinfo {author} {\bibfnamefont {K.}~\bibnamefont
  {Kang}}, \bibinfo {author} {\bibfnamefont {J.-H.}\ \bibnamefont {Park}},
  \bibinfo {author} {\bibfnamefont {H.}~\bibnamefont {Berger}}, \bibinfo
  {author} {\bibfnamefont {L.}~\bibnamefont {Forr{\'o}}}, \bibinfo {author}
  {\bibfnamefont {K.~T.}\ \bibnamefont {Law}},  \emph {et~al.},\ }\href
  {\doibase 10.1038/s41563-018-0061-1} {\bibfield  {journal} {\bibinfo
  {journal} {Nat. Mater.}\ }\textbf {\bibinfo {volume} {17}},\ \bibinfo {pages}
  {504} (\bibinfo {year} {2018})}\BibitemShut {NoStop}%
\bibitem [{\citenamefont {Barrera}\ \emph {et~al.}(2018)\citenamefont
  {Barrera}, \citenamefont {Sinko}, \citenamefont {Gopalan}, \citenamefont
  {Sivadas}, \citenamefont {Seyler}, \citenamefont {Watanabe}, \citenamefont
  {Taniguchi}, \citenamefont {Tsen}, \citenamefont {Xu}, \citenamefont {Xiao}
  \emph {et~al.}}]{Barrera2018}%
  \BibitemOpen
  \bibfield  {author} {\bibinfo {author} {\bibfnamefont {S.~C.}\ \bibnamefont
  {Barrera}}, \bibinfo {author} {\bibfnamefont {M.~R.}\ \bibnamefont {Sinko}},
  \bibinfo {author} {\bibfnamefont {D.~P.}\ \bibnamefont {Gopalan}}, \bibinfo
  {author} {\bibfnamefont {N.}~\bibnamefont {Sivadas}}, \bibinfo {author}
  {\bibfnamefont {K.~L.}\ \bibnamefont {Seyler}}, \bibinfo {author}
  {\bibfnamefont {K.}~\bibnamefont {Watanabe}}, \bibinfo {author}
  {\bibfnamefont {T.}~\bibnamefont {Taniguchi}}, \bibinfo {author}
  {\bibfnamefont {A.~W.}\ \bibnamefont {Tsen}}, \bibinfo {author}
  {\bibfnamefont {X.}~\bibnamefont {Xu}}, \bibinfo {author} {\bibfnamefont
  {D.}~\bibnamefont {Xiao}},  \emph {et~al.},\ }\href {\doibase
  10.1038/s41467-018-03888-4} {\bibfield  {journal} {\bibinfo  {journal} {Nat.
  Commun.}\ }\textbf {\bibinfo {volume} {9}},\ \bibinfo {pages} {1427}
  (\bibinfo {year} {2018})}\BibitemShut {NoStop}%
\bibitem [{\citenamefont {Yuan}\ \emph {et~al.}(2014)\citenamefont {Yuan},
  \citenamefont {Mak},\ and\ \citenamefont {Law}}]{Yuan2014}%
  \BibitemOpen
  \bibfield  {author} {\bibinfo {author} {\bibfnamefont {N.~F.~Q.}\
  \bibnamefont {Yuan}}, \bibinfo {author} {\bibfnamefont {K.~F.}\ \bibnamefont
  {Mak}}, \ and\ \bibinfo {author} {\bibfnamefont {K.~T.}\ \bibnamefont
  {Law}},\ }\href {\doibase 10.1103/PhysRevLett.113.097001} {\bibfield
  {journal} {\bibinfo  {journal} {Phys. Rev. Lett.}\ }\textbf {\bibinfo
  {volume} {113}},\ \bibinfo {pages} {097001} (\bibinfo {year}
  {2014})}\BibitemShut {NoStop}%
\bibitem [{\citenamefont {Zhou}\ \emph {et~al.}(2016)\citenamefont {Zhou},
  \citenamefont {Yuan}, \citenamefont {Jiang},\ and\ \citenamefont
  {Law}}]{Zhou2016}%
  \BibitemOpen
  \bibfield  {author} {\bibinfo {author} {\bibfnamefont {B.~T.}\ \bibnamefont
  {Zhou}}, \bibinfo {author} {\bibfnamefont {N.~F.~Q.}\ \bibnamefont {Yuan}},
  \bibinfo {author} {\bibfnamefont {H.-L.}\ \bibnamefont {Jiang}}, \ and\
  \bibinfo {author} {\bibfnamefont {K.~T.}\ \bibnamefont {Law}},\ }\href
  {\doibase 10.1103/PhysRevB.93.180501} {\bibfield  {journal} {\bibinfo
  {journal} {Phys. Rev. B}\ }\textbf {\bibinfo {volume} {93}},\ \bibinfo
  {pages} {180501} (\bibinfo {year} {2016})}\BibitemShut {NoStop}%
\bibitem [{\citenamefont {Sharma}\ and\ \citenamefont
  {Tewari}(2016)}]{Sharma2016}%
  \BibitemOpen
  \bibfield  {author} {\bibinfo {author} {\bibfnamefont {G.}~\bibnamefont
  {Sharma}}\ and\ \bibinfo {author} {\bibfnamefont {S.}~\bibnamefont
  {Tewari}},\ }\href {\doibase 10.1103/PhysRevB.94.094515} {\bibfield
  {journal} {\bibinfo  {journal} {Phys. Rev. B}\ }\textbf {\bibinfo {volume}
  {94}},\ \bibinfo {pages} {094515} (\bibinfo {year} {2016})}\BibitemShut
  {NoStop}%
\bibitem [{\citenamefont {Hsu}\ \emph {et~al.}(2017)\citenamefont {Hsu},
  \citenamefont {Vaezi}, \citenamefont {Fischer},\ and\ \citenamefont
  {Kim}}]{Hsu2017}%
  \BibitemOpen
  \bibfield  {author} {\bibinfo {author} {\bibfnamefont {Y.-T.}\ \bibnamefont
  {Hsu}}, \bibinfo {author} {\bibfnamefont {A.}~\bibnamefont {Vaezi}}, \bibinfo
  {author} {\bibfnamefont {M.~H.}\ \bibnamefont {Fischer}}, \ and\ \bibinfo
  {author} {\bibfnamefont {E.-A.}\ \bibnamefont {Kim}},\ }\href {\doibase
  10.1038/ncomms14985} {\bibfield  {journal} {\bibinfo  {journal} {Nat.
  Commun.}\ }\textbf {\bibinfo {volume} {8}},\ \bibinfo {pages} {14985}
  (\bibinfo {year} {2017})}\BibitemShut {NoStop}%
\bibitem [{\citenamefont {M\"ockli}\ and\ \citenamefont
  {Khodas}(2018)}]{Mockli2018}%
  \BibitemOpen
  \bibfield  {author} {\bibinfo {author} {\bibfnamefont {D.}~\bibnamefont
  {M\"ockli}}\ and\ \bibinfo {author} {\bibfnamefont {M.}~\bibnamefont
  {Khodas}},\ }\href {\doibase 10.1103/PhysRevB.98.144518} {\bibfield
  {journal} {\bibinfo  {journal} {Phys. Rev. B}\ }\textbf {\bibinfo {volume}
  {98}},\ \bibinfo {pages} {144518} (\bibinfo {year} {2018})}\BibitemShut
  {NoStop}%
\bibitem [{\citenamefont {Galvis}\ \emph {et~al.}(2014)\citenamefont {Galvis},
  \citenamefont {Chirolli}, \citenamefont {Guillam\'on}, \citenamefont
  {Vieira}, \citenamefont {Navarro-Moratalla}, \citenamefont {Coronado},
  \citenamefont {Suderow},\ and\ \citenamefont {Guinea}}]{Galvis2014}%
  \BibitemOpen
  \bibfield  {author} {\bibinfo {author} {\bibfnamefont {J.~A.}\ \bibnamefont
  {Galvis}}, \bibinfo {author} {\bibfnamefont {L.}~\bibnamefont {Chirolli}},
  \bibinfo {author} {\bibfnamefont {I.}~\bibnamefont {Guillam\'on}}, \bibinfo
  {author} {\bibfnamefont {S.}~\bibnamefont {Vieira}}, \bibinfo {author}
  {\bibfnamefont {E.}~\bibnamefont {Navarro-Moratalla}}, \bibinfo {author}
  {\bibfnamefont {E.}~\bibnamefont {Coronado}}, \bibinfo {author}
  {\bibfnamefont {H.}~\bibnamefont {Suderow}}, \ and\ \bibinfo {author}
  {\bibfnamefont {F.}~\bibnamefont {Guinea}},\ }\href {\doibase
  10.1103/PhysRevB.89.224512} {\bibfield  {journal} {\bibinfo  {journal} {Phys.
  Rev. B}\ }\textbf {\bibinfo {volume} {89}},\ \bibinfo {pages} {224512}
  (\bibinfo {year} {2014})}\BibitemShut {NoStop}%
\bibitem [{\citenamefont {Sanders}\ \emph {et~al.}(2016)\citenamefont
  {Sanders}, \citenamefont {Dendzik}, \citenamefont {Ngankeu}, \citenamefont
  {Eich}, \citenamefont {Bruix}, \citenamefont {Bianchi}, \citenamefont {Miwa},
  \citenamefont {Hammer}, \citenamefont {Khajetoorians},\ and\ \citenamefont
  {Hofmann}}]{Sanders2016}%
  \BibitemOpen
  \bibfield  {author} {\bibinfo {author} {\bibfnamefont {C.~E.}\ \bibnamefont
  {Sanders}}, \bibinfo {author} {\bibfnamefont {M.}~\bibnamefont {Dendzik}},
  \bibinfo {author} {\bibfnamefont {A.~S.}\ \bibnamefont {Ngankeu}}, \bibinfo
  {author} {\bibfnamefont {A.}~\bibnamefont {Eich}}, \bibinfo {author}
  {\bibfnamefont {A.}~\bibnamefont {Bruix}}, \bibinfo {author} {\bibfnamefont
  {M.}~\bibnamefont {Bianchi}}, \bibinfo {author} {\bibfnamefont {J.~A.}\
  \bibnamefont {Miwa}}, \bibinfo {author} {\bibfnamefont {B.}~\bibnamefont
  {Hammer}}, \bibinfo {author} {\bibfnamefont {A.~A.}\ \bibnamefont
  {Khajetoorians}}, \ and\ \bibinfo {author} {\bibfnamefont {P.}~\bibnamefont
  {Hofmann}},\ }\href {\doibase 10.1103/PhysRevB.94.081404} {\bibfield
  {journal} {\bibinfo  {journal} {Phys. Rev. B}\ }\textbf {\bibinfo {volume}
  {94}},\ \bibinfo {pages} {081404} (\bibinfo {year} {2016})}\BibitemShut
  {NoStop}%
\bibitem [{\citenamefont {Navarro-Moratalla}\ \emph {et~al.}(2016)\citenamefont
  {Navarro-Moratalla}, \citenamefont {Island}, \citenamefont
  {Ma{\~n}as-Valero}, \citenamefont {Pinilla-Cienfuegos}, \citenamefont
  {Castellanos-Gomez}, \citenamefont {Quereda}, \citenamefont
  {Rubio-Bollinger}, \citenamefont {Chirolli}, \citenamefont
  {Silva-Guill{\'e}n}, \citenamefont {Agra{\"\i}t} \emph
  {et~al.}}]{Navarro2016}%
  \BibitemOpen
  \bibfield  {author} {\bibinfo {author} {\bibfnamefont {E.}~\bibnamefont
  {Navarro-Moratalla}}, \bibinfo {author} {\bibfnamefont {J.}~\bibnamefont
  {Island}}, \bibinfo {author} {\bibfnamefont {S.}~\bibnamefont
  {Ma{\~n}as-Valero}}, \bibinfo {author} {\bibfnamefont {E.}~\bibnamefont
  {Pinilla-Cienfuegos}}, \bibinfo {author} {\bibfnamefont {A.}~\bibnamefont
  {Castellanos-Gomez}}, \bibinfo {author} {\bibfnamefont {J.}~\bibnamefont
  {Quereda}}, \bibinfo {author} {\bibfnamefont {G.}~\bibnamefont
  {Rubio-Bollinger}}, \bibinfo {author} {\bibfnamefont {L.}~\bibnamefont
  {Chirolli}}, \bibinfo {author} {\bibfnamefont {J.}~\bibnamefont
  {Silva-Guill{\'e}n}}, \bibinfo {author} {\bibfnamefont {N.}~\bibnamefont
  {Agra{\"\i}t}},  \emph {et~al.},\ }\href {\doibase 10.1038/ncomms11043}
  {\bibfield  {journal} {\bibinfo  {journal} {Nat. Commun.}\ }\textbf {\bibinfo
  {volume} {7}},\ \bibinfo {pages} {11043} (\bibinfo {year}
  {2016})}\BibitemShut {NoStop}%
\bibitem [{\citenamefont {Yang}\ \emph {et~al.}(2018)\citenamefont {Yang},
  \citenamefont {Fang}, \citenamefont {Fatemi}, \citenamefont {Ruhman},
  \citenamefont {Navarro-Moratalla}, \citenamefont {Watanabe}, \citenamefont
  {Taniguchi}, \citenamefont {Kaxiras},\ and\ \citenamefont
  {Jarillo-Herrero}}]{Yang2018}%
  \BibitemOpen
  \bibfield  {author} {\bibinfo {author} {\bibfnamefont {Y.}~\bibnamefont
  {Yang}}, \bibinfo {author} {\bibfnamefont {S.}~\bibnamefont {Fang}}, \bibinfo
  {author} {\bibfnamefont {V.}~\bibnamefont {Fatemi}}, \bibinfo {author}
  {\bibfnamefont {J.}~\bibnamefont {Ruhman}}, \bibinfo {author} {\bibfnamefont
  {E.}~\bibnamefont {Navarro-Moratalla}}, \bibinfo {author} {\bibfnamefont
  {K.}~\bibnamefont {Watanabe}}, \bibinfo {author} {\bibfnamefont
  {T.}~\bibnamefont {Taniguchi}}, \bibinfo {author} {\bibfnamefont
  {E.}~\bibnamefont {Kaxiras}}, \ and\ \bibinfo {author} {\bibfnamefont
  {P.}~\bibnamefont {Jarillo-Herrero}},\ }\href {\doibase
  10.1103/PhysRevB.98.035203} {\bibfield  {journal} {\bibinfo  {journal} {Phys.
  Rev. B}\ }\textbf {\bibinfo {volume} {98}},\ \bibinfo {pages} {035203}
  (\bibinfo {year} {2018})}\BibitemShut {NoStop}%
\bibitem [{\citenamefont {Dvir}\ \emph {et~al.}(2018)\citenamefont {Dvir},
  \citenamefont {Massee}, \citenamefont {Attias}, \citenamefont {Khodas},
  \citenamefont {Aprili}, \citenamefont {Quay},\ and\ \citenamefont
  {Steinberg}}]{Dvir2018}%
  \BibitemOpen
  \bibfield  {author} {\bibinfo {author} {\bibfnamefont {T.}~\bibnamefont
  {Dvir}}, \bibinfo {author} {\bibfnamefont {F.}~\bibnamefont {Massee}},
  \bibinfo {author} {\bibfnamefont {L.}~\bibnamefont {Attias}}, \bibinfo
  {author} {\bibfnamefont {M.}~\bibnamefont {Khodas}}, \bibinfo {author}
  {\bibfnamefont {M.}~\bibnamefont {Aprili}}, \bibinfo {author} {\bibfnamefont
  {C.~H.~L.}\ \bibnamefont {Quay}}, \ and\ \bibinfo {author} {\bibfnamefont
  {H.}~\bibnamefont {Steinberg}},\ }\href {\doibase 10.1038/s41467-018-03000-w}
  {\bibfield  {journal} {\bibinfo  {journal} {Nat. Commun.}\ }\textbf {\bibinfo
  {volume} {9}},\ \bibinfo {pages} {598} (\bibinfo {year} {2018})}\BibitemShut
  {NoStop}%
\bibitem [{\citenamefont {Ge}\ and\ \citenamefont {Liu}(2012)}]{Ge2012}%
  \BibitemOpen
  \bibfield  {author} {\bibinfo {author} {\bibfnamefont {Y.}~\bibnamefont
  {Ge}}\ and\ \bibinfo {author} {\bibfnamefont {A.~Y.}\ \bibnamefont {Liu}},\
  }\href {\doibase 10.1103/PhysRevB.86.104101} {\bibfield  {journal} {\bibinfo
  {journal} {Phys. Rev. B}\ }\textbf {\bibinfo {volume} {86}},\ \bibinfo
  {pages} {104101} (\bibinfo {year} {2012})}\BibitemShut {NoStop}%
\bibitem [{\citenamefont {Noat}\ \emph {et~al.}(2015)\citenamefont {Noat},
  \citenamefont {Silva-Guill\'en}, \citenamefont {Cren}, \citenamefont
  {Cherkez}, \citenamefont {Brun}, \citenamefont {Pons}, \citenamefont
  {Debontridder}, \citenamefont {Roditchev}, \citenamefont {Sacks},
  \citenamefont {Cario}, \citenamefont {Ordej\'on}, \citenamefont
  {Garc\'{\i}a},\ and\ \citenamefont {Canadell}}]{Noat2015}%
  \BibitemOpen
  \bibfield  {author} {\bibinfo {author} {\bibfnamefont {Y.}~\bibnamefont
  {Noat}}, \bibinfo {author} {\bibfnamefont {J.~A.}\ \bibnamefont
  {Silva-Guill\'en}}, \bibinfo {author} {\bibfnamefont {T.}~\bibnamefont
  {Cren}}, \bibinfo {author} {\bibfnamefont {V.}~\bibnamefont {Cherkez}},
  \bibinfo {author} {\bibfnamefont {C.}~\bibnamefont {Brun}}, \bibinfo {author}
  {\bibfnamefont {S.}~\bibnamefont {Pons}}, \bibinfo {author} {\bibfnamefont
  {F.}~\bibnamefont {Debontridder}}, \bibinfo {author} {\bibfnamefont
  {D.}~\bibnamefont {Roditchev}}, \bibinfo {author} {\bibfnamefont
  {W.}~\bibnamefont {Sacks}}, \bibinfo {author} {\bibfnamefont
  {L.}~\bibnamefont {Cario}}, \bibinfo {author} {\bibfnamefont
  {P.}~\bibnamefont {Ordej\'on}}, \bibinfo {author} {\bibfnamefont
  {A.}~\bibnamefont {Garc\'{\i}a}}, \ and\ \bibinfo {author} {\bibfnamefont
  {E.}~\bibnamefont {Canadell}},\ }\href {\doibase 10.1103/PhysRevB.92.134510}
  {\bibfield  {journal} {\bibinfo  {journal} {Phys. Rev. B}\ }\textbf {\bibinfo
  {volume} {92}},\ \bibinfo {pages} {134510} (\bibinfo {year}
  {2015})}\BibitemShut {NoStop}%
\bibitem [{\citenamefont {Heil}\ \emph {et~al.}(2017)\citenamefont {Heil},
  \citenamefont {Ponc\'e}, \citenamefont {Lambert}, \citenamefont {Schlipf},
  \citenamefont {Margine},\ and\ \citenamefont {Giustino}}]{Heil2017}%
  \BibitemOpen
  \bibfield  {author} {\bibinfo {author} {\bibfnamefont {C.}~\bibnamefont
  {Heil}}, \bibinfo {author} {\bibfnamefont {S.}~\bibnamefont {Ponc\'e}},
  \bibinfo {author} {\bibfnamefont {H.}~\bibnamefont {Lambert}}, \bibinfo
  {author} {\bibfnamefont {M.}~\bibnamefont {Schlipf}}, \bibinfo {author}
  {\bibfnamefont {E.~R.}\ \bibnamefont {Margine}}, \ and\ \bibinfo {author}
  {\bibfnamefont {F.}~\bibnamefont {Giustino}},\ }\href {\doibase
  10.1103/PhysRevLett.119.087003} {\bibfield  {journal} {\bibinfo  {journal}
  {Phys. Rev. Lett.}\ }\textbf {\bibinfo {volume} {119}},\ \bibinfo {pages}
  {087003} (\bibinfo {year} {2017})}\BibitemShut {NoStop}%
\bibitem [{\citenamefont {Zhao}\ \emph {et~al.}(2017)\citenamefont {Zhao},
  \citenamefont {Wijayaratne}, \citenamefont {Butler}, \citenamefont {Yang},
  \citenamefont {Malliakas}, \citenamefont {Chung}, \citenamefont {Louca},
  \citenamefont {Kanatzidis}, \citenamefont {van Wezel},\ and\ \citenamefont
  {Chatterjee}}]{Zhao2017}%
  \BibitemOpen
  \bibfield  {author} {\bibinfo {author} {\bibfnamefont {J.}~\bibnamefont
  {Zhao}}, \bibinfo {author} {\bibfnamefont {K.}~\bibnamefont {Wijayaratne}},
  \bibinfo {author} {\bibfnamefont {A.}~\bibnamefont {Butler}}, \bibinfo
  {author} {\bibfnamefont {J.}~\bibnamefont {Yang}}, \bibinfo {author}
  {\bibfnamefont {C.~D.}\ \bibnamefont {Malliakas}}, \bibinfo {author}
  {\bibfnamefont {D.~Y.}\ \bibnamefont {Chung}}, \bibinfo {author}
  {\bibfnamefont {D.}~\bibnamefont {Louca}}, \bibinfo {author} {\bibfnamefont
  {M.~G.}\ \bibnamefont {Kanatzidis}}, \bibinfo {author} {\bibfnamefont
  {J.}~\bibnamefont {van Wezel}}, \ and\ \bibinfo {author} {\bibfnamefont
  {U.}~\bibnamefont {Chatterjee}},\ }\href {\doibase
  10.1103/PhysRevB.96.125103} {\bibfield  {journal} {\bibinfo  {journal} {Phys.
  Rev. B}\ }\textbf {\bibinfo {volume} {96}},\ \bibinfo {pages} {125103}
  (\bibinfo {year} {2017})}\BibitemShut {NoStop}%
\bibitem [{\citenamefont {Youn}\ \emph {et~al.}(2012)\citenamefont {Youn},
  \citenamefont {Fischer}, \citenamefont {Rhim}, \citenamefont {Sigrist},\ and\
  \citenamefont {Agterberg}}]{Youn2012}%
  \BibitemOpen
  \bibfield  {author} {\bibinfo {author} {\bibfnamefont {S.~J.}\ \bibnamefont
  {Youn}}, \bibinfo {author} {\bibfnamefont {M.~H.}\ \bibnamefont {Fischer}},
  \bibinfo {author} {\bibfnamefont {S.~H.}\ \bibnamefont {Rhim}}, \bibinfo
  {author} {\bibfnamefont {M.}~\bibnamefont {Sigrist}}, \ and\ \bibinfo
  {author} {\bibfnamefont {D.~F.}\ \bibnamefont {Agterberg}},\ }\href {\doibase
  10.1103/PhysRevB.85.220505} {\bibfield  {journal} {\bibinfo  {journal} {Phys.
  Rev. B}\ }\textbf {\bibinfo {volume} {85}},\ \bibinfo {pages} {220505}
  (\bibinfo {year} {2012})}\BibitemShut {NoStop}%
\bibitem [{\citenamefont {Liu}(2017)}]{Liu2017}%
  \BibitemOpen
  \bibfield  {author} {\bibinfo {author} {\bibfnamefont {C.-X.}\ \bibnamefont
  {Liu}},\ }\href {\doibase 10.1103/PhysRevLett.118.087001} {\bibfield
  {journal} {\bibinfo  {journal} {Phys. Rev. Lett.}\ }\textbf {\bibinfo
  {volume} {118}},\ \bibinfo {pages} {087001} (\bibinfo {year}
  {2017})}\BibitemShut {NoStop}%
\bibitem [{\citenamefont {Mizukami}\ \emph {et~al.}(2011)\citenamefont
  {Mizukami}, \citenamefont {Shishido}, \citenamefont {Shibauchi},
  \citenamefont {Shimozawa}, \citenamefont {Yasumoto}, \citenamefont
  {Watanabe}, \citenamefont {Yamashita}, \citenamefont {Ikeda}, \citenamefont
  {Terashima}, \citenamefont {Kontani} \emph {et~al.}}]{Mizukami2011}%
  \BibitemOpen
  \bibfield  {author} {\bibinfo {author} {\bibfnamefont {Y.}~\bibnamefont
  {Mizukami}}, \bibinfo {author} {\bibfnamefont {H.}~\bibnamefont {Shishido}},
  \bibinfo {author} {\bibfnamefont {T.}~\bibnamefont {Shibauchi}}, \bibinfo
  {author} {\bibfnamefont {M.}~\bibnamefont {Shimozawa}}, \bibinfo {author}
  {\bibfnamefont {S.}~\bibnamefont {Yasumoto}}, \bibinfo {author}
  {\bibfnamefont {D.}~\bibnamefont {Watanabe}}, \bibinfo {author}
  {\bibfnamefont {M.}~\bibnamefont {Yamashita}}, \bibinfo {author}
  {\bibfnamefont {H.}~\bibnamefont {Ikeda}}, \bibinfo {author} {\bibfnamefont
  {T.}~\bibnamefont {Terashima}}, \bibinfo {author} {\bibfnamefont
  {H.}~\bibnamefont {Kontani}},  \emph {et~al.},\ }\href {\doibase
  10.1038/nphys2112} {\bibfield  {journal} {\bibinfo  {journal} {Nat. Phys.}\
  }\textbf {\bibinfo {volume} {7}},\ \bibinfo {pages} {849} (\bibinfo {year}
  {2011})}\BibitemShut {NoStop}%
\bibitem [{\citenamefont {Goh}\ \emph {et~al.}(2012)\citenamefont {Goh},
  \citenamefont {Mizukami}, \citenamefont {Shishido}, \citenamefont {Watanabe},
  \citenamefont {Yasumoto}, \citenamefont {Shimozawa}, \citenamefont
  {Yamashita}, \citenamefont {Terashima}, \citenamefont {Yanase}, \citenamefont
  {Shibauchi}, \citenamefont {Buzdin},\ and\ \citenamefont
  {Matsuda}}]{Goh2012}%
  \BibitemOpen
  \bibfield  {author} {\bibinfo {author} {\bibfnamefont {S.~K.}\ \bibnamefont
  {Goh}}, \bibinfo {author} {\bibfnamefont {Y.}~\bibnamefont {Mizukami}},
  \bibinfo {author} {\bibfnamefont {H.}~\bibnamefont {Shishido}}, \bibinfo
  {author} {\bibfnamefont {D.}~\bibnamefont {Watanabe}}, \bibinfo {author}
  {\bibfnamefont {S.}~\bibnamefont {Yasumoto}}, \bibinfo {author}
  {\bibfnamefont {M.}~\bibnamefont {Shimozawa}}, \bibinfo {author}
  {\bibfnamefont {M.}~\bibnamefont {Yamashita}}, \bibinfo {author}
  {\bibfnamefont {T.}~\bibnamefont {Terashima}}, \bibinfo {author}
  {\bibfnamefont {Y.}~\bibnamefont {Yanase}}, \bibinfo {author} {\bibfnamefont
  {T.}~\bibnamefont {Shibauchi}}, \bibinfo {author} {\bibfnamefont {A.~I.}\
  \bibnamefont {Buzdin}}, \ and\ \bibinfo {author} {\bibfnamefont
  {Y.}~\bibnamefont {Matsuda}},\ }\href {\doibase
  10.1103/PhysRevLett.109.157006} {\bibfield  {journal} {\bibinfo  {journal}
  {Phys. Rev. Lett.}\ }\textbf {\bibinfo {volume} {109}},\ \bibinfo {pages}
  {157006} (\bibinfo {year} {2012})}\BibitemShut {NoStop}%
\bibitem [{\citenamefont {Shimozawa}\ \emph {et~al.}(2014)\citenamefont
  {Shimozawa}, \citenamefont {Goh}, \citenamefont {Endo}, \citenamefont
  {Kobayashi}, \citenamefont {Watashige}, \citenamefont {Mizukami},
  \citenamefont {Ikeda}, \citenamefont {Shishido}, \citenamefont {Yanase},
  \citenamefont {Terashima}, \citenamefont {Shibauchi},\ and\ \citenamefont
  {Matsuda}}]{Shimozawa2014}%
  \BibitemOpen
  \bibfield  {author} {\bibinfo {author} {\bibfnamefont {M.}~\bibnamefont
  {Shimozawa}}, \bibinfo {author} {\bibfnamefont {S.~K.}\ \bibnamefont {Goh}},
  \bibinfo {author} {\bibfnamefont {R.}~\bibnamefont {Endo}}, \bibinfo {author}
  {\bibfnamefont {R.}~\bibnamefont {Kobayashi}}, \bibinfo {author}
  {\bibfnamefont {T.}~\bibnamefont {Watashige}}, \bibinfo {author}
  {\bibfnamefont {Y.}~\bibnamefont {Mizukami}}, \bibinfo {author}
  {\bibfnamefont {H.}~\bibnamefont {Ikeda}}, \bibinfo {author} {\bibfnamefont
  {H.}~\bibnamefont {Shishido}}, \bibinfo {author} {\bibfnamefont
  {Y.}~\bibnamefont {Yanase}}, \bibinfo {author} {\bibfnamefont
  {T.}~\bibnamefont {Terashima}}, \bibinfo {author} {\bibfnamefont
  {T.}~\bibnamefont {Shibauchi}}, \ and\ \bibinfo {author} {\bibfnamefont
  {Y.}~\bibnamefont {Matsuda}},\ }\href {\doibase
  10.1103/PhysRevLett.112.156404} {\bibfield  {journal} {\bibinfo  {journal}
  {Phys. Rev. Lett.}\ }\textbf {\bibinfo {volume} {112}},\ \bibinfo {pages}
  {156404} (\bibinfo {year} {2014})}\BibitemShut {NoStop}%
\bibitem [{\citenamefont {Sigrist}\ and\ \citenamefont
  {Ueda}(1991)}]{Sigrist1991}%
  \BibitemOpen
  \bibfield  {author} {\bibinfo {author} {\bibfnamefont {M.}~\bibnamefont
  {Sigrist}}\ and\ \bibinfo {author} {\bibfnamefont {K.}~\bibnamefont {Ueda}},\
  }\href {\doibase 10.1103/RevModPhys.63.239} {\bibfield  {journal} {\bibinfo
  {journal} {Rev. Mod. Phys.}\ }\textbf {\bibinfo {volume} {63}},\ \bibinfo
  {pages} {239} (\bibinfo {year} {1991})}\BibitemShut {NoStop}%
\bibitem [{\citenamefont {Yoshida}\ \emph {et~al.}(2015)\citenamefont
  {Yoshida}, \citenamefont {Sigrist},\ and\ \citenamefont
  {Yanase}}]{Yoshida2015}%
  \BibitemOpen
  \bibfield  {author} {\bibinfo {author} {\bibfnamefont {T.}~\bibnamefont
  {Yoshida}}, \bibinfo {author} {\bibfnamefont {M.}~\bibnamefont {Sigrist}}, \
  and\ \bibinfo {author} {\bibfnamefont {Y.}~\bibnamefont {Yanase}},\ }\href
  {\doibase 10.1103/PhysRevLett.115.027001} {\bibfield  {journal} {\bibinfo
  {journal} {Phys. Rev. Lett.}\ }\textbf {\bibinfo {volume} {115}},\ \bibinfo
  {pages} {027001} (\bibinfo {year} {2015})}\BibitemShut {NoStop}%
\bibitem [{\citenamefont {Fang}\ \emph {et~al.}(2005)\citenamefont {Fang},
  \citenamefont {Wang}, \citenamefont {Zou}, \citenamefont {Tang},
  \citenamefont {Xu}, \citenamefont {Chen}, \citenamefont {Dong}, \citenamefont
  {Shan},\ and\ \citenamefont {Wen}}]{Fang2005}%
  \BibitemOpen
  \bibfield  {author} {\bibinfo {author} {\bibfnamefont {L.}~\bibnamefont
  {Fang}}, \bibinfo {author} {\bibfnamefont {Y.}~\bibnamefont {Wang}}, \bibinfo
  {author} {\bibfnamefont {P.~Y.}\ \bibnamefont {Zou}}, \bibinfo {author}
  {\bibfnamefont {L.}~\bibnamefont {Tang}}, \bibinfo {author} {\bibfnamefont
  {Z.}~\bibnamefont {Xu}}, \bibinfo {author} {\bibfnamefont {H.}~\bibnamefont
  {Chen}}, \bibinfo {author} {\bibfnamefont {C.}~\bibnamefont {Dong}}, \bibinfo
  {author} {\bibfnamefont {L.}~\bibnamefont {Shan}}, \ and\ \bibinfo {author}
  {\bibfnamefont {H.~H.}\ \bibnamefont {Wen}},\ }\href {\doibase
  10.1103/PhysRevB.72.014534} {\bibfield  {journal} {\bibinfo  {journal} {Phys.
  Rev. B}\ }\textbf {\bibinfo {volume} {72}},\ \bibinfo {pages} {014534}
  (\bibinfo {year} {2005})}\BibitemShut {NoStop}%
\bibitem [{\citenamefont {Wagner}\ \emph {et~al.}(2008)\citenamefont {Wagner},
  \citenamefont {Morosan}, \citenamefont {Hor}, \citenamefont {Tao},
  \citenamefont {Zhu}, \citenamefont {Sanders}, \citenamefont {McQueen},
  \citenamefont {Zandbergen}, \citenamefont {Williams}, \citenamefont {West},\
  and\ \citenamefont {Cava}}]{Wagner2008}%
  \BibitemOpen
  \bibfield  {author} {\bibinfo {author} {\bibfnamefont {K.~E.}\ \bibnamefont
  {Wagner}}, \bibinfo {author} {\bibfnamefont {E.}~\bibnamefont {Morosan}},
  \bibinfo {author} {\bibfnamefont {Y.~S.}\ \bibnamefont {Hor}}, \bibinfo
  {author} {\bibfnamefont {J.}~\bibnamefont {Tao}}, \bibinfo {author}
  {\bibfnamefont {Y.}~\bibnamefont {Zhu}}, \bibinfo {author} {\bibfnamefont
  {T.}~\bibnamefont {Sanders}}, \bibinfo {author} {\bibfnamefont {T.~M.}\
  \bibnamefont {McQueen}}, \bibinfo {author} {\bibfnamefont {H.~W.}\
  \bibnamefont {Zandbergen}}, \bibinfo {author} {\bibfnamefont {A.~J.}\
  \bibnamefont {Williams}}, \bibinfo {author} {\bibfnamefont {D.~V.}\
  \bibnamefont {West}}, \ and\ \bibinfo {author} {\bibfnamefont {R.~J.}\
  \bibnamefont {Cava}},\ }\href {\doibase 10.1103/PhysRevB.78.104520}
  {\bibfield  {journal} {\bibinfo  {journal} {Phys. Rev. B}\ }\textbf {\bibinfo
  {volume} {78}},\ \bibinfo {pages} {104520} (\bibinfo {year}
  {2008})}\BibitemShut {NoStop}%
\bibitem [{\citenamefont {Albertini}\ \emph {et~al.}(2017)\citenamefont
  {Albertini}, \citenamefont {Liu},\ and\ \citenamefont
  {Calandra}}]{Albertini2017}%
  \BibitemOpen
  \bibfield  {author} {\bibinfo {author} {\bibfnamefont {O.~R.}\ \bibnamefont
  {Albertini}}, \bibinfo {author} {\bibfnamefont {A.~Y.}\ \bibnamefont {Liu}},
  \ and\ \bibinfo {author} {\bibfnamefont {M.}~\bibnamefont {Calandra}},\
  }\href {\doibase 10.1103/PhysRevB.95.235121} {\bibfield  {journal} {\bibinfo
  {journal} {Phys. Rev. B}\ }\textbf {\bibinfo {volume} {95}},\ \bibinfo
  {pages} {235121} (\bibinfo {year} {2017})}\BibitemShut {NoStop}%
\bibitem [{\citenamefont {Lian}\ \emph {et~al.}(2017)\citenamefont {Lian},
  \citenamefont {Si}, \citenamefont {Wu},\ and\ \citenamefont
  {Duan}}]{Lian2017}%
  \BibitemOpen
  \bibfield  {author} {\bibinfo {author} {\bibfnamefont {C.-S.}\ \bibnamefont
  {Lian}}, \bibinfo {author} {\bibfnamefont {C.}~\bibnamefont {Si}}, \bibinfo
  {author} {\bibfnamefont {J.}~\bibnamefont {Wu}}, \ and\ \bibinfo {author}
  {\bibfnamefont {W.}~\bibnamefont {Duan}},\ }\href {\doibase
  10.1103/PhysRevB.96.235426} {\bibfield  {journal} {\bibinfo  {journal} {Phys.
  Rev. B}\ }\textbf {\bibinfo {volume} {96}},\ \bibinfo {pages} {235426}
  (\bibinfo {year} {2017})}\BibitemShut {NoStop}%
\bibitem [{\citenamefont {Xi}\ \emph {et~al.}(2016)\citenamefont {Xi},
  \citenamefont {Berger}, \citenamefont {Forr\'o}, \citenamefont {Shan},\ and\
  \citenamefont {Mak}}]{Xi2016}%
  \BibitemOpen
  \bibfield  {author} {\bibinfo {author} {\bibfnamefont {X.}~\bibnamefont
  {Xi}}, \bibinfo {author} {\bibfnamefont {H.}~\bibnamefont {Berger}}, \bibinfo
  {author} {\bibfnamefont {L.}~\bibnamefont {Forr\'o}}, \bibinfo {author}
  {\bibfnamefont {J.}~\bibnamefont {Shan}}, \ and\ \bibinfo {author}
  {\bibfnamefont {K.~F.}\ \bibnamefont {Mak}},\ }\href {\doibase
  10.1103/PhysRevLett.117.106801} {\bibfield  {journal} {\bibinfo  {journal}
  {Phys. Rev. Lett.}\ }\textbf {\bibinfo {volume} {117}},\ \bibinfo {pages}
  {106801} (\bibinfo {year} {2016})}\BibitemShut {NoStop}%
\bibitem [{\citenamefont {Suderow}\ \emph {et~al.}(2005)\citenamefont
  {Suderow}, \citenamefont {Tissen}, \citenamefont {Brison}, \citenamefont
  {Mart\'{\i}nez},\ and\ \citenamefont {Vieira}}]{Suderow2005}%
  \BibitemOpen
  \bibfield  {author} {\bibinfo {author} {\bibfnamefont {H.}~\bibnamefont
  {Suderow}}, \bibinfo {author} {\bibfnamefont {V.~G.}\ \bibnamefont {Tissen}},
  \bibinfo {author} {\bibfnamefont {J.~P.}\ \bibnamefont {Brison}}, \bibinfo
  {author} {\bibfnamefont {J.~L.}\ \bibnamefont {Mart\'{\i}nez}}, \ and\
  \bibinfo {author} {\bibfnamefont {S.}~\bibnamefont {Vieira}},\ }\href
  {\doibase 10.1103/PhysRevLett.95.117006} {\bibfield  {journal} {\bibinfo
  {journal} {Phys. Rev. Lett.}\ }\textbf {\bibinfo {volume} {95}},\ \bibinfo
  {pages} {117006} (\bibinfo {year} {2005})}\BibitemShut {NoStop}%
\bibitem [{\citenamefont {Tissen}\ \emph {et~al.}(2013)\citenamefont {Tissen},
  \citenamefont {Osorio}, \citenamefont {Brison}, \citenamefont {Nemes},
  \citenamefont {Garc\'{\i}a-Hern\'andez}, \citenamefont {Cario}, \citenamefont
  {Rodi\`ere}, \citenamefont {Vieira},\ and\ \citenamefont
  {Suderow}}]{Tissen2013}%
  \BibitemOpen
  \bibfield  {author} {\bibinfo {author} {\bibfnamefont {V.~G.}\ \bibnamefont
  {Tissen}}, \bibinfo {author} {\bibfnamefont {M.~R.}\ \bibnamefont {Osorio}},
  \bibinfo {author} {\bibfnamefont {J.~P.}\ \bibnamefont {Brison}}, \bibinfo
  {author} {\bibfnamefont {N.~M.}\ \bibnamefont {Nemes}}, \bibinfo {author}
  {\bibfnamefont {M.}~\bibnamefont {Garc\'{\i}a-Hern\'andez}}, \bibinfo
  {author} {\bibfnamefont {L.}~\bibnamefont {Cario}}, \bibinfo {author}
  {\bibfnamefont {P.}~\bibnamefont {Rodi\`ere}}, \bibinfo {author}
  {\bibfnamefont {S.}~\bibnamefont {Vieira}}, \ and\ \bibinfo {author}
  {\bibfnamefont {H.}~\bibnamefont {Suderow}},\ }\href {\doibase
  10.1103/PhysRevB.87.134502} {\bibfield  {journal} {\bibinfo  {journal} {Phys.
  Rev. B}\ }\textbf {\bibinfo {volume} {87}},\ \bibinfo {pages} {134502}
  (\bibinfo {year} {2013})}\BibitemShut {NoStop}%
\bibitem [{\citenamefont {Freitas}\ \emph {et~al.}(2016)\citenamefont
  {Freitas}, \citenamefont {Rodi\`ere}, \citenamefont {Osorio}, \citenamefont
  {Navarro-Moratalla}, \citenamefont {Nemes}, \citenamefont {Tissen},
  \citenamefont {Cario}, \citenamefont {Coronado}, \citenamefont
  {Garc\'{\i}a-Hern\'andez}, \citenamefont {Vieira}, \citenamefont {N\'u\~nez
  Regueiro},\ and\ \citenamefont {Suderow}}]{Freitas2016}%
  \BibitemOpen
  \bibfield  {author} {\bibinfo {author} {\bibfnamefont {D.~C.}\ \bibnamefont
  {Freitas}}, \bibinfo {author} {\bibfnamefont {P.}~\bibnamefont {Rodi\`ere}},
  \bibinfo {author} {\bibfnamefont {M.~R.}\ \bibnamefont {Osorio}}, \bibinfo
  {author} {\bibfnamefont {E.}~\bibnamefont {Navarro-Moratalla}}, \bibinfo
  {author} {\bibfnamefont {N.~M.}\ \bibnamefont {Nemes}}, \bibinfo {author}
  {\bibfnamefont {V.~G.}\ \bibnamefont {Tissen}}, \bibinfo {author}
  {\bibfnamefont {L.}~\bibnamefont {Cario}}, \bibinfo {author} {\bibfnamefont
  {E.}~\bibnamefont {Coronado}}, \bibinfo {author} {\bibfnamefont
  {M.}~\bibnamefont {Garc\'{\i}a-Hern\'andez}}, \bibinfo {author}
  {\bibfnamefont {S.}~\bibnamefont {Vieira}}, \bibinfo {author} {\bibfnamefont
  {M.}~\bibnamefont {N\'u\~nez Regueiro}}, \ and\ \bibinfo {author}
  {\bibfnamefont {H.}~\bibnamefont {Suderow}},\ }\href {\doibase
  10.1103/PhysRevB.93.184512} {\bibfield  {journal} {\bibinfo  {journal} {Phys.
  Rev. B}\ }\textbf {\bibinfo {volume} {93}},\ \bibinfo {pages} {184512}
  (\bibinfo {year} {2016})}\BibitemShut {NoStop}%
\bibitem [{\citenamefont {Grasset}\ \emph {et~al.}(2018)\citenamefont
  {Grasset}, \citenamefont {Gallais}, \citenamefont {Sacuto}, \citenamefont
  {Cazayous}, \citenamefont {Ma{\~n}as-Valero}, \citenamefont {Coronado},\ and\
  \citenamefont {M{\'e}asson}}]{Grasset2018}%
  \BibitemOpen
  \bibfield  {author} {\bibinfo {author} {\bibfnamefont {R.}~\bibnamefont
  {Grasset}}, \bibinfo {author} {\bibfnamefont {Y.}~\bibnamefont {Gallais}},
  \bibinfo {author} {\bibfnamefont {A.}~\bibnamefont {Sacuto}}, \bibinfo
  {author} {\bibfnamefont {M.}~\bibnamefont {Cazayous}}, \bibinfo {author}
  {\bibfnamefont {S.}~\bibnamefont {Ma{\~n}as-Valero}}, \bibinfo {author}
  {\bibfnamefont {E.}~\bibnamefont {Coronado}}, \ and\ \bibinfo {author}
  {\bibfnamefont {M.-A.}\ \bibnamefont {M{\'e}asson}},\ }\href@noop {}
  {\bibfield  {journal} {\bibinfo  {journal} {arXiv:1806.03433}\ } (\bibinfo
  {year} {2018})}\BibitemShut {NoStop}%
\end{thebibliography}%
	
\end{document}